\documentclass{optica-article}
\usepackage[utf8]{inputenc}
\journal{opticajournal} 
\articletype{Research Article}
\usepackage{lineno}
\usepackage{dcolumn}
\usepackage{braket}
\usepackage{mathtools}
\usepackage{graphicx}
\usepackage{mathrsfs}
\usepackage{mdwlist}
\usepackage{subfigure}
\usepackage{booktabs}
\usepackage{amsmath}
\usepackage{appendix}
\usepackage{extarrows}
\usepackage{dsfont}
\usepackage{amstext}
\usepackage{amsbsy}
\usepackage{bbm}
\usepackage{bm}
\usepackage{graphicx}
\usepackage{textcomp}
\usepackage{multirow}
\usepackage{threeparttable}

\usepackage{color}
\usepackage{diagbox}
\definecolor{Dgreen}{RGB}{0, 100, 0}
\usepackage{url}

\begin{document}
	\title{Spontaneous emission in Casimir-Rabi oscillations through a weak optomechanical coupling}
	\author{Ke-Xiong Yan,\authormark{1,2} Yuan Qiu,\authormark{1,2} Yang Xiao,\authormark{1,2}  Jie Song,\authormark{5} Ye-Hong Chen,\authormark{1,2,3,4,6} and \author{Yan Xia,\authormark{1,2,7}}}
	
	\address{\authormark{1}Fujian Key Laboratory of Quantum Information and Quantum Optics (Fuzhou University), Fuzhou 350116, China\\
		\authormark{2}Department of Physics, Fuzhou University, Fuzhou 350116, China\\
		\authormark{3}Theoretical Quantum Physics Laboratory, Cluster for Pioneering Research, RIKEN, Wako-shi, Saitama 351-0198, Japan\\
		\authormark{4}Quantum Information Physics Theory Research Team, Center for Quantum Computing, RIKEN, Wako-shi, Saitama 351-0198, Japan\\
		\authormark{5}School of Physics, Harbin Institute of Technology, Harbin 150001, China\\
\authormark{6}yehong.chen@fzu.edu.cn	\\
\authormark{7}xia-208@163.com}

		\begin{abstract}
		
         When the mechanical oscillator is initially in an excited state, its free evolution, driven by the dynamical Casimir effect (DCE), leads to the generation of radiation in the vacuum, analogous to the spontaneous emission of an excited atom.  In the regime where the loss rates are lower than the corresponding frequency splittings, reversible energy exchange can occur between the excited mechanical oscillator and the vacuum cavity field (i.e. vacuum Casimir-Rabi oscillations). By using the quantum trajectory approach, we show that $n$-photon/$n$-phonon emission can be achieved by opening  dissipative channels for these Casimir-Rabi oscillations. Simulation results from $10^{5}$ quantum trajectories demonstrate that the purities of two-photon emission, two-phonon emission, and three-phonon emission can reach 0.92, 0.94, and 0.95, respectively. These values are observed in Casimir-Rabi oscillations between the three-phonon state and the two-photon state in a weak optomechanical coupling. This phenomenon could provide insights into the physical characteristics of photon and phonon emission in the DCE.
	\end{abstract}

	\section{INTRODUCTION}
	Quantum field theory predicts that the quantum vacuum is not empty, but teeming with virtual particles~\cite{PhysRev.82.664,Moore1970,HAWKING1974,PhysRevLett.62.1742,Schwinger1993,Wilson2011,RevModPhys.84.1,Lhteenmki2013}. By applying certain external perturbations, these vacuum fluctuations can be converted into real particles. Examples include the Schwinger effect~\cite{PhysRev.82.664}, Hawking radiation~\cite{HAWKING1974}, Unruh effect~\cite{PhysRevD.14.870}, and dynamical Casimir effect (DCE)~\cite{Barton1993,Wilson2011,RevModPhys.84.1,Lhteenmki2013,1976Fulling}. As one of the most representative examples, the DCE describes the generation of photons due to rapid changes in the geometry (in particular the positions of some boundaries) or material properties of electrically neutral macroscopic or mesoscopic objects. Although DCE was predicted in 1976~\cite{Moore1970}, radiation induced by vibrating mirrors has no yet been observed. One of the main difficulties is that the speed of vibration of mirrors can hardly approach the speed of light~\cite{Dodonov2010,RevModPhys.84.1}.
	
	In order to circumvent these difficulties, a number of alternative proposals have been put forth. These theoretical proposals have suggested that the mechanical motion can be replaced with an effective one~\cite{PhysRevLett.93.193601,SEGEV2007202,PhysRevLett.110.243601,Chen2024,PhysRevA.110.043711}, which may be provided by, for example, modulating the light-matter coupling strength in cavity quantum electrodynamics (QED) systems~\cite{PhysRevB.93.235309,Baker2017,PhysRevLett.126.023602,PhysRevLett.119.053601,RevModPhys.91.025005,PhysRevLett.131.113601,PhysRevLett.133.033603,QIN20241}, modulating the boundary condition of a superconducting waveguide or resonator~\cite{Wilson2011,PhysRevLett.103.147003,PhysRevA.82.052509}, and using lasers to rapidly modulate the reflectivity of thin semiconductor films~\cite{PhysRevA.70.033811,Braggio2005}. In particular, two noteworthy experimental confirmations have been attained through the utilization of a superconducting quantum interference device~\cite{RevModPhys.84.1,PhysRevLett.103.147003,PhysRevA.82.052509,Wilson2011,Dalvit2011,PhysRevA.87.043804} and Josephson metamaterials~\cite{Lhteenmki2013} to generate an effective motion, respectively. In Ref.~\cite{PhysRevX.8.011031}, \text{ Macrì} $et~al.$ have found that vacuum radiation can originate from the free evolution of a pure mechanical excited state, which provides a more fundamental explanation of the DCE. They have shown that a resonant production of photons out from the vacuum can be observed even when the mechanical frequency $\omega_{m}$ is lower than the cavity-mode frequency $\omega_{c}$ in the cavity-optomechanical system. This regime can mitigate the problem mentioned above (i.e. very fast oscillating mirror). 
	
	In this manuscript, we extend the framework proposed by \text{ Macrì} $et~al.$~\cite{PhysRevX.8.011031} by investigating n-photon/n-phonon emission in Casimir-Rabi oscillations between the three-phonon state and the two-photon state under weak optomechanical coupling. Such oscillations can be generated under the condition where the mechanical frequency is lower than the optical frequency. This makes it a particularly promising example for experimental investigation. Unlike conventional Rabi oscillations driven by linear light-matter interactions, the DCE-induced coupling enables multiphoton-multiphonon exchange processes, governed by a third-order effective Hamiltonian. We present a comprehensive theoretical analysis of the virtual pathways involved in the transition process and rigorously derive the high-order effective Hamiltonian to accurately calculate the resonance condition. Subsequently, we simulate the spontaneous emission process of Casimir-Rabi oscillations  through the stochastic evolution of the system wave function~\cite{PhysRevLett.52.1657,Disi1986,PhysRevA.46.4363,PhysRevLett.56.2797}. The quantum trajectory approach, i.e. the stochastic evolution of the wave function, allows for the investigation and quantification of the emission of photons and phonons in each DCE simulation. The simulation result indicates that the number of trajectories responsible for radiation generation can become significant when the mechanical dissipation rate is lower than the photon dissipation rate. Furthermore, we observe the occurrence of two-photon and two- or three-phonon emission in photon and phonon emission processes, respectively. When the coupling strength is $0.05\omega_{m}$ and both the phonon and photon dissipation rates are $10^{-9}\omega_{m}$, the purities of two-photon emission, two-phonon emission, and three-phonon emission are 0.92, 0.94, and 0.95, respectively. This pattern may assist in developing a deeper comprehension of the physical characteristics of photon and phonon emission in the DCE.
	
	The rest of the paper is organized as follows: In Sec.~\ref{s2}, we introduce the model and calculate the effective Hamiltonian. In Sec.~\ref{s3}, the quantum trajectory approach is employed for the investigation of spontaneous emission from the mechanically excited state. This section explores $n$-photon ($n$-phonon) emission processes, the effects of temperature on the emission of photons (phonons), and the system's sensitivity to various parameters. The potential experimental protocol is presented in Sec.~\ref{s4}. Finally, the work is summarized in Sec.~\ref{s6}.
	
	\section{THE MODEL AND EFFECTIVE HAMILTONIAN} \label{s2}
    
\begin{figure}
		\centering
		\includegraphics[scale=0.5]{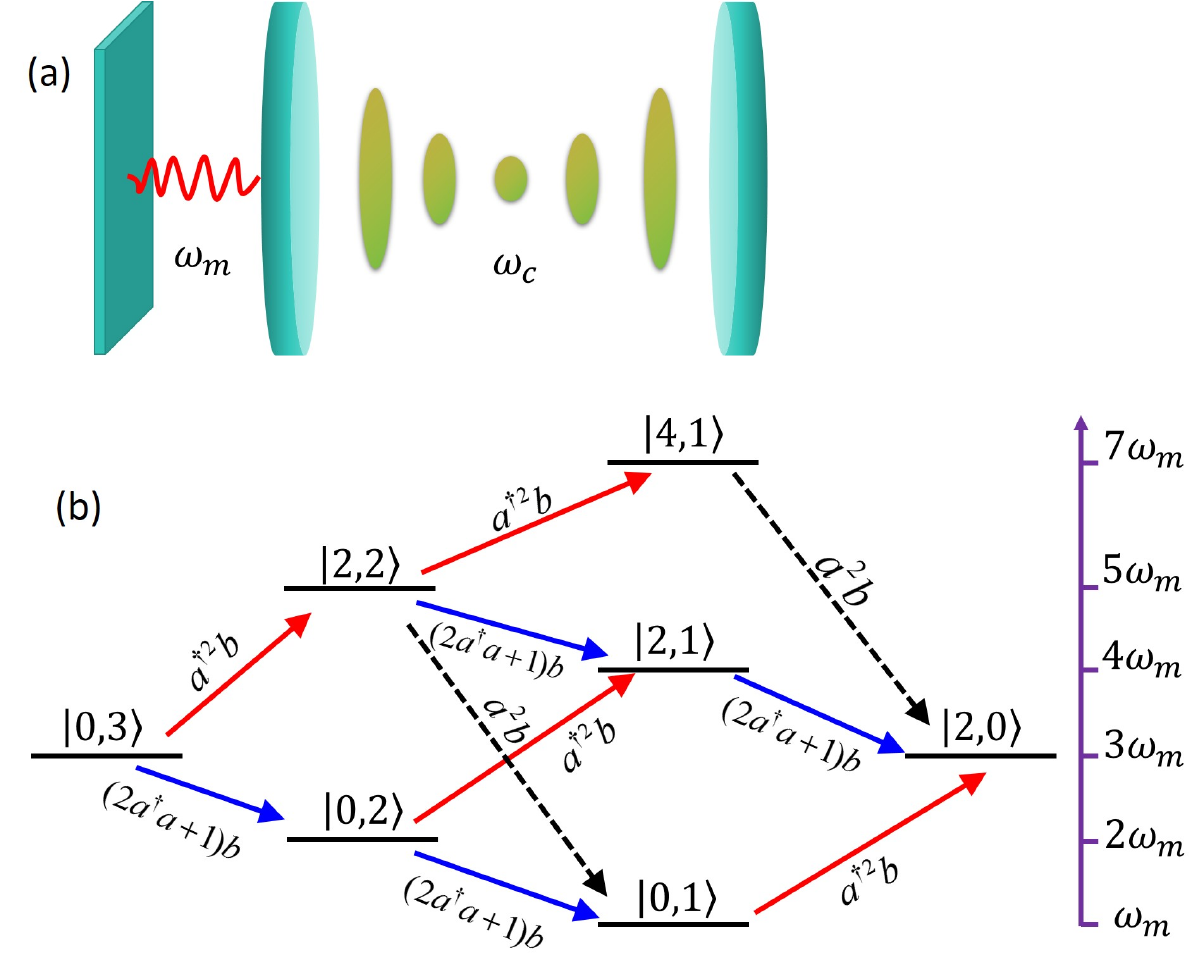}
		\caption{(a) Schematic of a typical optomechanical system where one of the mirrors in an optical cavity can vibrate at frequency $\omega_{m}$. The resonance frequency of the cavity is $\omega_{c}$. (b) All the possible virtual paths for the transition $\ket{0,3}\rightarrow \ket{2,0}$. }
		\label{F1}
    \end{figure}

	We consider the case of an optical cavity with a movable end mirror, as schematically depicted in Fig.~\ref{F1}(a). The Hamiltonian of this system can be written as 
	\begin{eqnarray}
    H_{s}=\omega_{c}a^{\dagger}a+\omega_{m}b^{\dagger}b+g(a^{\dagger}+a)^{2}(b^{\dagger}+b),
		\label{eq1}
	\end{eqnarray}
	where $a$ ($a^{\dagger}$) is the annihilation (creation) operator of the optical mode of the cavity with the lowest-frequency $\omega_{c}$. The resonance frequency of the mechanical mode is $\omega_{m}$ with bosonic operators $b$ and $b^{\dagger}$. The optomechanical coupling rate is $g$. A detailed derivation of this Hamiltonian can be found in Ref.~\cite{PhysRevA.51.2537}.
    \begin{figure}
	\centering
	\includegraphics[scale=0.5]{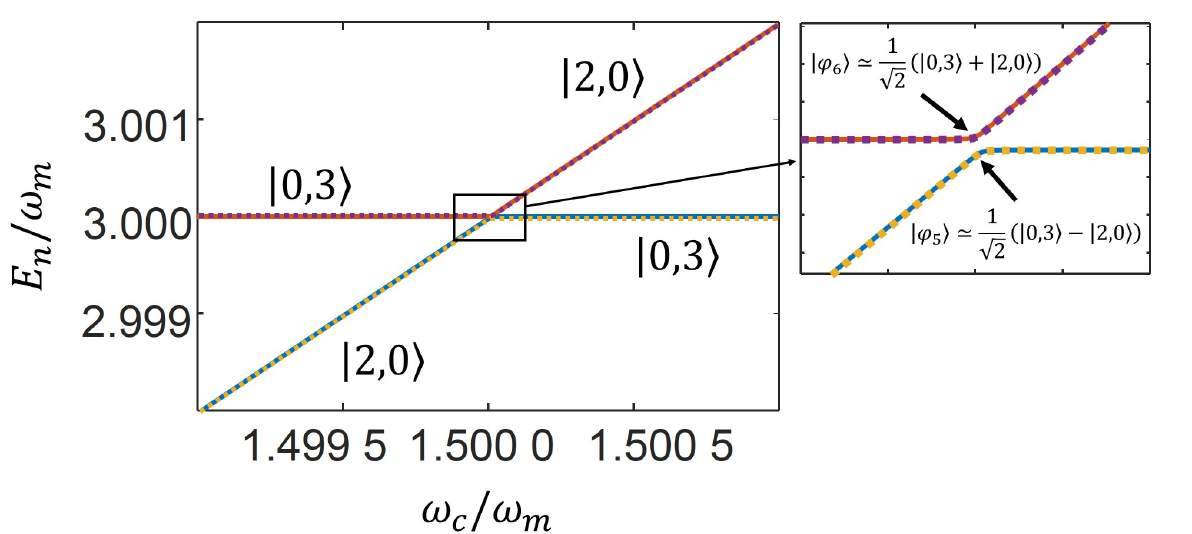}
	\caption{Eigenvalues $E_{5}/ \omega_{m}$ and $E_{6}/ \omega_{m}$ as a function of the ratio between the cavity frequency $\omega_{c}$ and the mechanical frequency $\omega_{m}$. The continuous curves are the eigenvalues of $H_{s}$ in Eq.~(\ref{eq1}) and the dashed lines describe the eigenenergies of the $H_{\rm eff}$ in Eq.~(\ref{eq3}). When energy level splitting is minimal, $\omega_{c}=1.5000105\omega_{m}$. Other parameter is $g=0.001\omega_{m}$.}
	\label{F2}
\end{figure} 
	By numerically diagonalizing the Hamiltonian $H_{s}$ in Eq.~(\ref{eq1}), we can obtain the energy spectrum of the system.  The energy spectrum exhibits numerous avoided crossings~\cite{PhysRevX.8.011031}.  These avoided crossings result from the coherent coupling between the state $\ket{n,k}$ and the state $\ket{n+2,k-q}$ with $k\geq q$. The state $\ket{n,k}=\ket{n}_{c}\otimes\ket{k}_{m}$, where Fock states $\ket{n}_{c}$ and $\ket{k}_{m}$ are the eigenstates of the optical and mechanical modes, respectively. The variables $n$, $k$, and $q$ denote the number of photon excitations, the number of phonon excitations, and the number of annihilated phonons, respectively. 
    
    Here, we investigate the coherent coupling between the three-phonon state $\ket{0,3}$ and the two-photon state $\ket{2,0}$. The coupling between these two states ensures that the mechanical frequency is lower than the optical frequency, while maintaining a large effective coupling strength. In Fig.~\ref{F1}(b), we present all possible effective energy level transition processes. The system Hamiltonian, given by Eq.~(\ref{eq1}), commutes with the photon parity operator, i.e., $[H_{s},e^{i\pi a^{\dagger}a}]=0$. This implies that the intermediate states in each transition path must satisfy the condition of even photon parity. Moreover, since the interaction terms in Eq.~(\ref{eq1}) involve only $(a^{\dagger})^2b$, $a^2 b$, $(a^{\dagger}a+1)b$, and their conjugate terms, each step in the effective transition path must adhere the following transition rules: $\Delta n_{c}=0,\pm2$ and $\Delta n_{p}=\pm1$, where $\Delta n_{c}$ and $\Delta n_{p}$ denote the changes in the photon and phonon numbers, respectively. We only focus on the fifth excited state $\ket{\varphi_{5}}$ and the sixth excited state $\ket{\varphi_{6}}$ (that is, $H_{s}\ket{\varphi_{n}}=E_{n}\ket{\varphi_{n}}$ with $n=5, 6$), and the corresponding energy levels $E_{5}$ and $E_{6}$ as a function of the ratio between the cavity and the mechanical frequency are presented in Fig.~\ref{F2}. It can be found that the two energy levels exhibit an avoided crossing, which serves as a signature of a particular closed subdynamics. When the vacuum Casimir-Rabi splitting~\cite{PhysRevX.8.011031} is at its minimum (i.e. $\omega_{c}\simeq 3\omega_{m}/2$), the two system eigenstates are approximately the symmetric and antisymmetric superposition states
	\begin{eqnarray}
		\ket{\varphi_{5(6)}}\simeq \frac{1}{\sqrt{2}}(\ket{0,3}\pm \ket{2,0}).
	\end{eqnarray} 
    
    When the detuning $\Delta=\omega_{c}-\omega_{m}\simeq1/3\omega_{c}\gg g$, an effective Hamiltonian can be derived using the generalized James' effective Hamiltonian method~\cite{PhysRevA.95.032124} (see Appendix~\ref{app:a}):
    \begin{eqnarray}
    	\nonumber
    	H_{\rm eff}&=&\omega_{c}a^{\dagger}a+\omega_{m}b^{\dagger}b+H_{\rm eff}^{(2)}+H_{\rm eff}^{(3)},\\\nonumber
    	H_{\rm eff}^{(2)}&=&\frac{g^{2}}{4\omega_{m}}\left[a^{\dagger2}a^{2}-2(2a^{\dagger}a+1)(3b^{\dagger}b+4a^{\dagger}a+3)\right],\\
    	H_{\rm eff}^{(3)}&=&\frac{9g^{3}}{\omega_{m}^{2}}(a^{\dagger2}b^{3}+a^{2}b^{\dagger3}).
    	\label{eq3}
    \end{eqnarray}

     The second-order effective Hamiltonian $H_{\rm eff}^{(2)}$ leads to a shift in the resonance frequency and the third-order effective Hamiltonian $H_{\rm eff}^{(3)}$ is the one responsible for the Casimir-Rabi oscillation between the state $\ket{0,3}$ and the state $\ket{2,0}$. In the subspace formed by $\ket{0,3}$ and $\ket{2,0}$, the matrix form of the effective Hamiltonian $H_{{\rm eff}}$ can be written as
     \begin{equation}
     	\setlength{\arraycolsep}{8pt}
     	H_{{\rm eff}}=\begin{pmatrix}
     		3\omega_{m}-\frac{6g^{2}}{\omega_{m}} & 
     		\frac{18\sqrt{3}g^{3}}{\omega_{m}^{2}} \\[2ex]
     		\frac{18\sqrt{3}g^{3}}{\omega_{m}^{2}} &
     		2\omega_{c}-\frac{27g^{2}}{\omega_{m}} \label{eq44}
     	\end{pmatrix}.
     \end{equation}
     The aforementioned matrix allows us to ascertain the effective Rabi frequency $\Omega_{\rm eff}$ and the resonance condition $\omega_{c}$:
     \begin{eqnarray}
     	\nonumber
     	\frac{\Omega_{\rm eff}}{\omega_{m}}=\frac{18\sqrt{3}g^{3}}{\omega_{m}^{3}},\ \ \ 
     	\frac{\omega_{c}}{\omega_{m}}=\frac{3}{2}+\frac{21g^{2}}{2\omega_{m}^2}.
     \end{eqnarray}
     
     Actually, the effective Hamiltonian in Eq.~(\ref{eq3}) describes the transition process between the state $\ket{n,k}$ and the state $\ket{n+2,k-3}$ (where $n$ and $k$ are the number of photon excitations and phonon excitations, respectively). In the subspace formed by these two states, we can calculate the corresponding Rabi frequency and resonance condition as follows:
     \begin{eqnarray}
     	\frac{\Omega_{\rm eff}}{\omega_{m}}&=&\frac{9\sqrt{(n+1)(n+2)k(k-1)(k-2)}g^{3}}{\omega_{m}^{3}},\\
     	\frac{\omega_{c}}{\omega_{m}}&=&\frac{3}{2}+\frac{3g^{2}(2n+2k+1)}{2\omega_{m}^2}.
     \end{eqnarray}
     Due to the sensitivity of this high-order process, even a small resonance frequency shift (induced by the second-order effective Hamiltonian) can suppress the transition channel. Similar transition processes involving additional photons or phonons ($k>3$ or $n>0$), where the difference between their corresponding resonance frequencies and those of the subspace of interest is much greater than the effective coupling strength, can be neglected.   For example, consider the same kind of transition process $\ket{1,3}\rightarrow \ket{3,0}$ with a resonance frequency $\omega_{c}=1.5000135\omega_{m}$. The resonance frequency of this similar transition process differs significantly from that of the transition process $\ket{0,3}\rightarrow \ket{2,0}$ we are interested in by $\delta \omega=3\times10^{-6}\omega_{m}$, which is much larger than the effective coupling strength $\Omega_{\rm eff}=18\sqrt{3}\times10^{-9}\omega_{m}$. Therefore, this similar transition process does not contribute in the final process and can be neglected. 

     \begin{figure}
     	\centering
     	\includegraphics[scale=0.6]{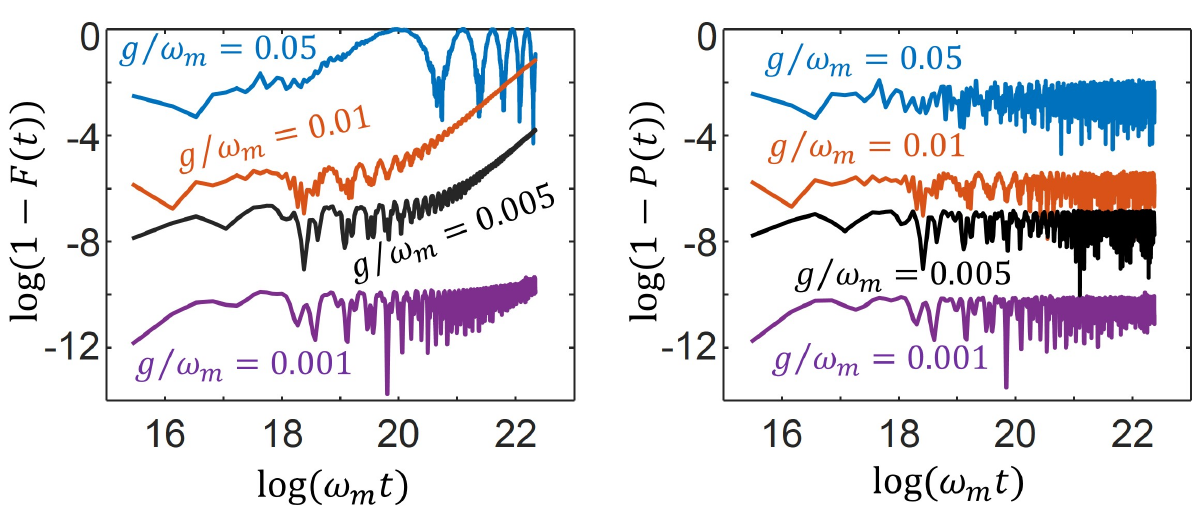}
     	\caption{(a) Fidelity $F(t)$ defined in Eq.~(\ref{eq4}) 
 (b) Population $P(t)$ vs the time for various values of the coupling rate: $g=0.001\omega_{m}$ (purple solid curve), $g=0.005\omega_{m}$ (black solid curve),  $g=0.01\omega_{m}$ (orange solid curve), and $g=0.05$ (blue solid curve). Other parameter is $\omega_{c}=1.5000105\omega_{m}$. The initial state of the system is $\ket{0,3}$. }
     	\label{F3}
     \end{figure}
     In the derivation of Eq.~(\ref{eq3}), the rotating-wave approximation (RWA) is employed, necessitating an evaluation of its validity. The validity of the RWA can be ascertained by examining the fidelity
     \begin{equation}
     	F(t)=|\langle \phi(t)|\psi(t)\rangle|^{2},
     	\label{eq4}
     \end{equation}
     between the state $\ket{\phi(t)}$ obtained by solving the Schr\"{o}dinger equation with the exact Hamiltonian~(\ref{eq1}) and the state $\ket{\psi(t)}$ obtained by solving the Schr\"{o}dinger equation with the effective Hamiltonian~(\ref{eq3}). Figure~\ref{F3}(a) shows that the logarithm of infidelity $1-F(t)$ undergoes oscillations over time, with an increase observed as the coupling strength decreases. When the coupling strength is $g=0.001\omega_{m}$,  $1-F(t)<10^{-8}$, which provides evidence that the RWA is valid. Furthermore, we plot the population $P(t)$ in the subspace spanned $\{ \ket{0,3}, \ket{2,0}\}$ with time. The definition of population is given by 
     \begin{equation}
         P(t)=|\langle 0,3|\phi(t)\rangle|^{2}+|\langle 2,0|\phi(t)\rangle|^{2}.
     \end{equation}
     From Fig.~\ref{F3}(b), it can be found that when the coupling strength reaches $g/\omega_{m}=0.05$, the population that leaks outside the subspace remains small. This indicates that oscillations between the states $\ket{0,3}$ and $\ket{2,0}$ are also possible only when the resonance condition is precisely satisfied. It should be emphasized that when the coupling strength gradually increases, the effective Hamiltonian in Eq.~(\ref{eq44}) cannot accurately determine the resonance conditions for three-phonon-two-photon oscillations. This is why the population leaking outside the subspace is very small (around $10^{-4}$), but the logarithmic value of infidelity is still very large and oscillates over time.

	\section{SPONTANEOUS EMISSION OF THE MECHANICALLY EXCITED STATE} \label{s3}
	Spontaneous emission is a process by which a quantum emitter (such as atom and molecule) decays from an excited state to a lower energy state and emits photons~\cite{PhysRevA.44.657}. It is a remarkable manifestation of the interaction between a quantum system and vacuum fields. In this section, we discuss the spontaneous emission of the mechanically excited state by following individual trajectories of the evolution of the system.
	
	\begin{figure}
		\centering
        \subfigure{
			\label{F4(a)}
			\includegraphics[scale=0.41]{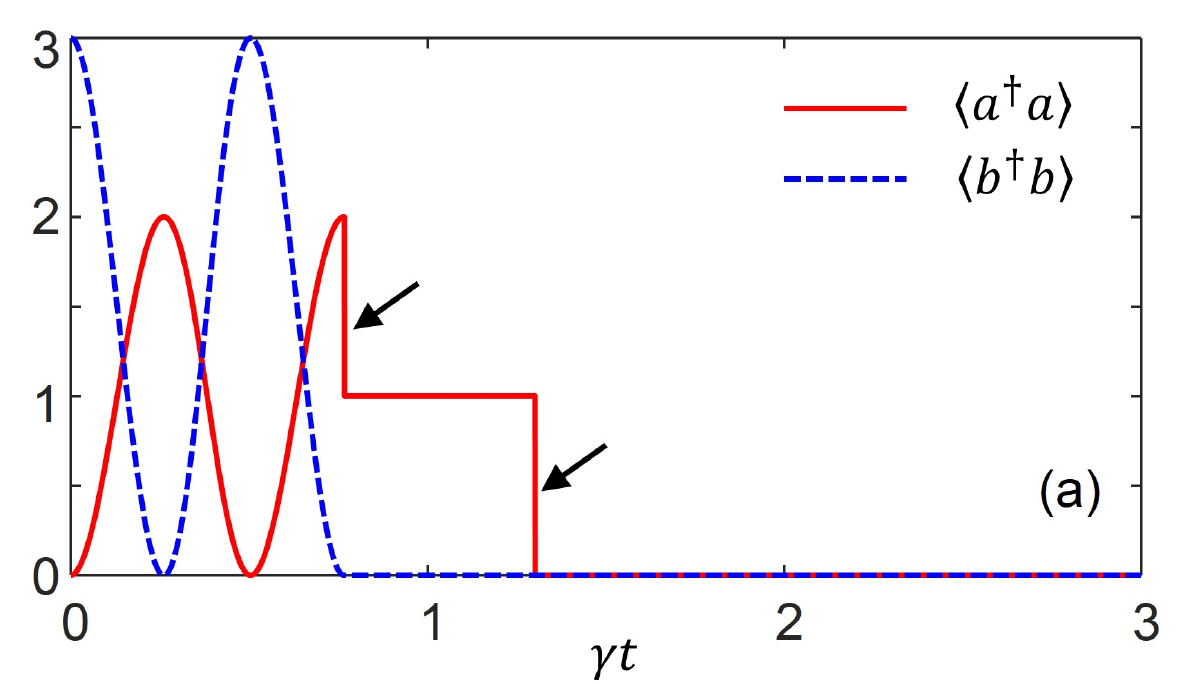}
            \label{F4a}
		}
	\subfigure{
			\label{F4(b)}
			\includegraphics[scale=0.41]{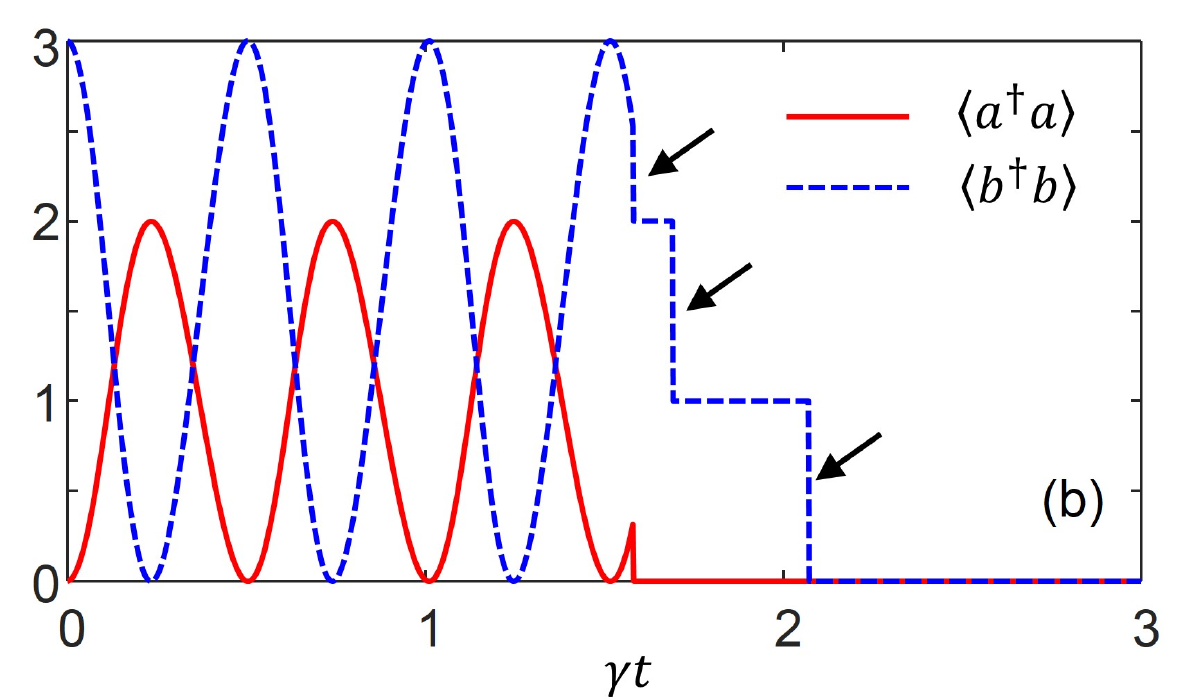}
            \label{F4b}
		}
		\caption{Examples of a single quantum trajectory, numerically obtained by studying the open quantum dynamics. It shows the time evolution of the mean excitation number of the radiation field $\langle a^{\dagger}a \rangle $~(red solid curve) and of the mechanical oscillator $ \langle b^{\dagger}b  \rangle $~(blue dotted curve). The black arrows in both panels indicate that the system has undergone a quantum jump. In panel~(a), the system emits two photons outward with an emission interval less than the photon lifetime. In panel~(b), the system emits three phonons outward with an emission interval less than the phonon lifetime. In both panels, the system is initialized in $\ket{0,3}$ and other parameters are $g=0.001\omega_{m}$, $\omega_{c}=1.5000105\omega_{m}$, and $\gamma=\gamma_{a}=\gamma_{b}=10^{-9}\omega_{m}$.}
		\label{F4}
	\end{figure} 

    \begin{figure}
		\centering
        \includegraphics[scale=0.65]{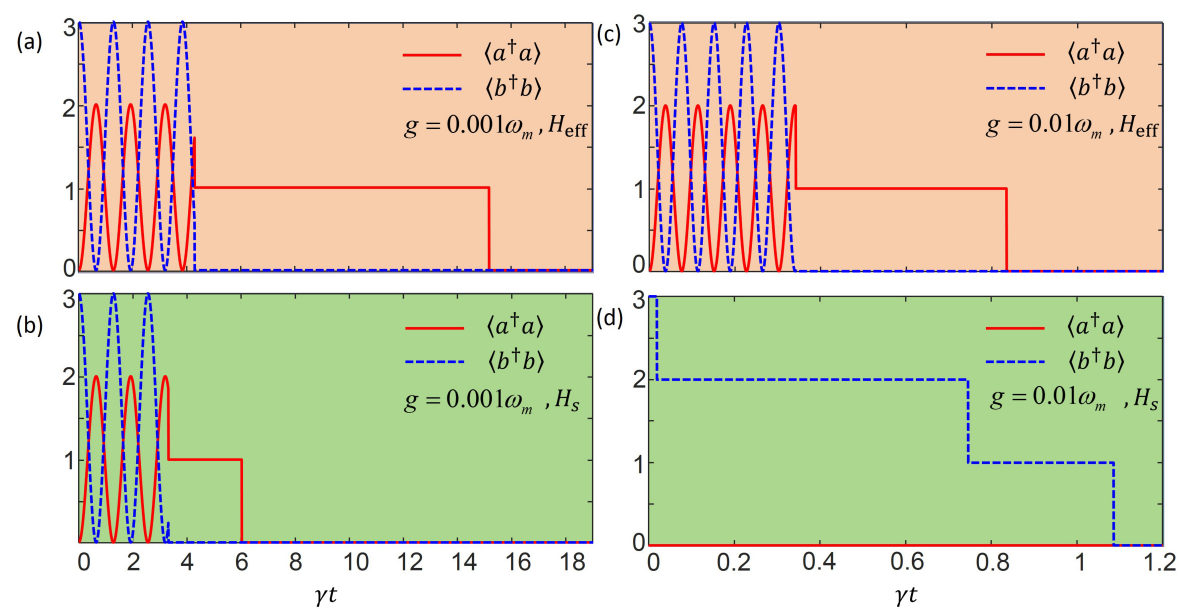}
	 \caption{(a) and (c) show a single quantum trajectory obtained from the effective Hamiltonian in Eq.~(\ref{eq3}) at different coupling strengths, respectively. The numerical result in (b) and (d) are obtained from the system Hamiltonian in Eq.~(\ref{eq1}). In both panels, the system is initialized in $\ket{0,3}$ and other parameters are $\omega_{c}=1.5000105\omega_{m}$ and $\gamma=\gamma_{a}=\gamma_{b}=10^{-9}\omega_{m}$. }
		\label{F44}
	\end{figure} 
    
	\subsection{Quantum trajectory}
	 Dissipation of energy from the system to environment is typically treated by a density operator approach, which is also known as a master equation approach. This approach is capable of describing the ensemble average over numerous trajectories of a quantum system, but requires substantial computational resources~\cite{PhysRevLett.68.580}.
	 By solving the stochastic evolution of the system wave function~\cite{PhysRevLett.68.580,Molmer:93,PhysRevLett.124.053601,PhysRevResearch.5.013221}, one can follow the individual trajectories of the system and record the clicks of photons (phonons). In order to describe the effect of the environment between two quantum jumps, we introduce the non-Hermitian Hamiltonian
	\begin{equation}
		\mathcal{H}=H_{s}-i(\gamma_{a}a^{\dagger}a+\gamma_{b}b^{\dagger}b)/2,
	\end{equation}
	where $H_{s}$ represents the system Hamiltonian in Eq.~(\ref{eq1}), $\gamma_{a}$ and $\gamma_{b}$ are the photonic and mechanical loss rates, respectively. The evolution of a quantum trajectory is thus dictated by a non-Hermitian evolution via $H_{s}$ interrupted by random quantum jumps. The algorithm to obtain such a dynamics can be found in refs~\cite{PhysRevLett.68.580,Molmer:93,PhysRevLett.124.053601,PhysRevResearch.5.013221,PhysRevA.105.023720}.
	
	In the subspace spanned by $\{\ket{0,3}, \ket{2,0}\}$, we can have a simple analytical description. If we initialize the system in the state $\ket{0,3}$, the system state at time $t$, before a quantum jump takes place, is 
	\begin{equation}
		\ket{\psi(t)}=e^{-\gamma t/2}[\cos (\Omega_{\rm eff}t)\ket{0,3}-i\sin (\Omega_{\rm eff}t)\ket{2,0}],
	\end{equation}
    where we choose $\gamma=\gamma_{a}=\gamma_{b}$.
    After the system wave function is normalized, the evolution equations for the average phonon number and photon number over time can be obtained:
    \begin{eqnarray}
    	\nonumber
    	\left \langle a^{\dagger}a\right \rangle &=&2\sin^{2} (\Omega_{\rm eff}t)\\ 
    	\left \langle b^{\dagger}b\right \rangle &=&3\cos^{2} (\Omega_{\rm eff}t).
    	\label{eq8}
    \end{eqnarray}
	
	The numerical results are shown in Fig.~\ref{F4}. We note that in Fig.~\ref{F4a} the system initially oscillates between the state $\ket{0,3}$ and the state $\ket{2,0}$ predicted by Eq.~(\ref{eq8}). When a quantum jump occurs, one photon has escaped from the cavity and the state of system collapses to a single photon state $\ket{1,0}$. This state interrupts the Casimir-Rabi oscillation between the state $\ket{0,3}$ and the state $\ket{2,0}$ and is preserved until a second jump occurs. After the emission of the second photon in the cavity, the system is in the vacuum state $\ket{0,0}$. Throughout the whole process, the system emits only two photons, no phonons are emitted. This is because before the first quantum jump, the system is almost in a two-photon state $\ket{2,0}$ with only photon excitation and no phonon excitation. After the quantum jump occurs, the oscillation between the state $\ket{0,3}$ and the state $\ket{2,0}$ is interrupted, making the conversion between photon and phonon impossible, and there is no more phonon excitation. Similar dynamics process are presented in Fig.~\ref{F4b}. 

    In Figs.~\ref{F44}, we present the difference in simulation results obtained from the effective Hamiltonian in Eq.~(\ref{eq3}) and the total Hamiltonian in Eq.~(\ref{eq1}), respectively. As the coupling strength increases, trajectories under the effective Hamiltonian maintain coherent oscillations between $\ket{0,3}$ and $\ket{2,0}$, and the oscillation frequency increases with the coupling strength. However, trajectories evolving under the full Hamiltonian in Eq.~(\ref{eq1}) exhibit complete suppression of oscillations, yielding exclusively continuous phonon emission. This difference arises because, in quantum trajectory simulations using the effective Hamiltonian, the evolution of the system's dynamics is confined to the subspace spanned by the $\ket{0,3}$ and $\ket{2,0}$ states, which allows three-phonon-two-photon oscillations to occur even when the resonance conditions are not perfectly satisfied. In contrast, when simulations are performed using the system Hamiltonian, deviations from the resonance conditions cause the system's dynamics to be restricted to other subspaces, thereby preventing the evolution of the three-phonon state. This phenomenon indicates that the validity of RWA rapidly decreases with increasing coupling strength, further corroborating the numerical results in Fig.~\ref{F3}. In the following discussion, we will consistently use the system Hamiltonian in Eq.~(\ref{eq1}) for trajectory simulations, with numerical calculations employed to ensure that the resonance conditions are accurately satisfied for different coupling strengths.

\begin{figure}
	\centering
	\includegraphics[scale=0.55]{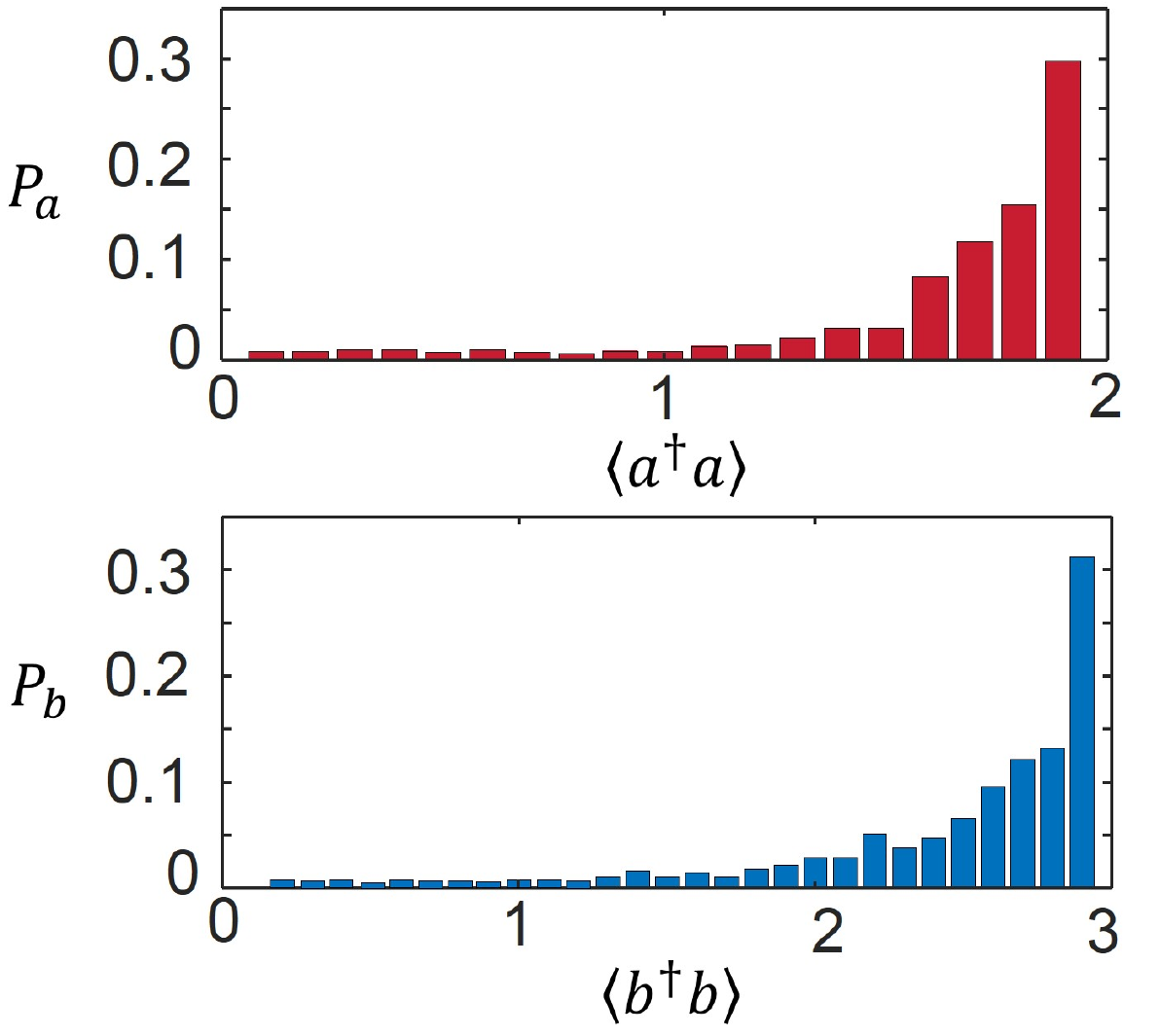}
	\caption{Histogram of the ratio of photon (phonon) jumps as a function of the average photon (phonon) number.  The histogram is constructed from simulations of  $10^{5}$ trajectories, consisting of 40,529 trajectories for photon emission and 59,471 trajectories for phonon emission.  All parameters are chosen as in Fig.~\ref{F4}.}
	\label{F5}
\end{figure}

\begin{figure}
	\centering
	\includegraphics[scale=0.55]{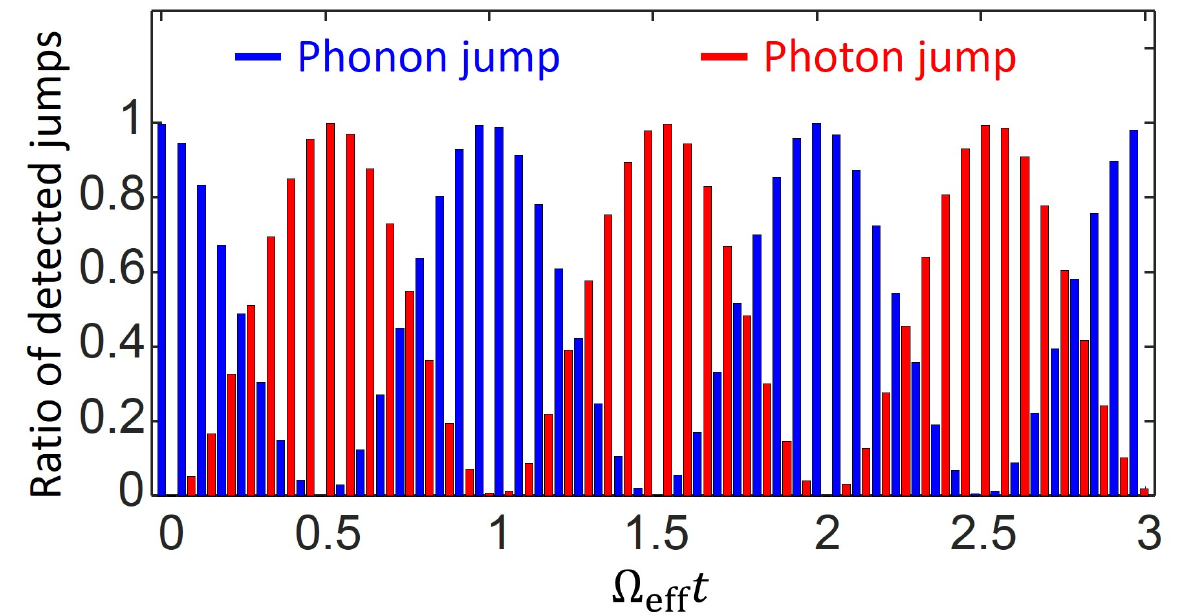}
	\caption{Histograms of the ratio of the total quantum jumps as a function of time. The histogram is constructed from simulations of $10^{5}$ trajectories. The system is initialized in the state $\ket{0,3}$ . All parameters are chosen as in Fig.~\ref{F4}.}
	\label{F12}
\end{figure}
\begin{figure}
		\centering
		\includegraphics[scale=0.55]{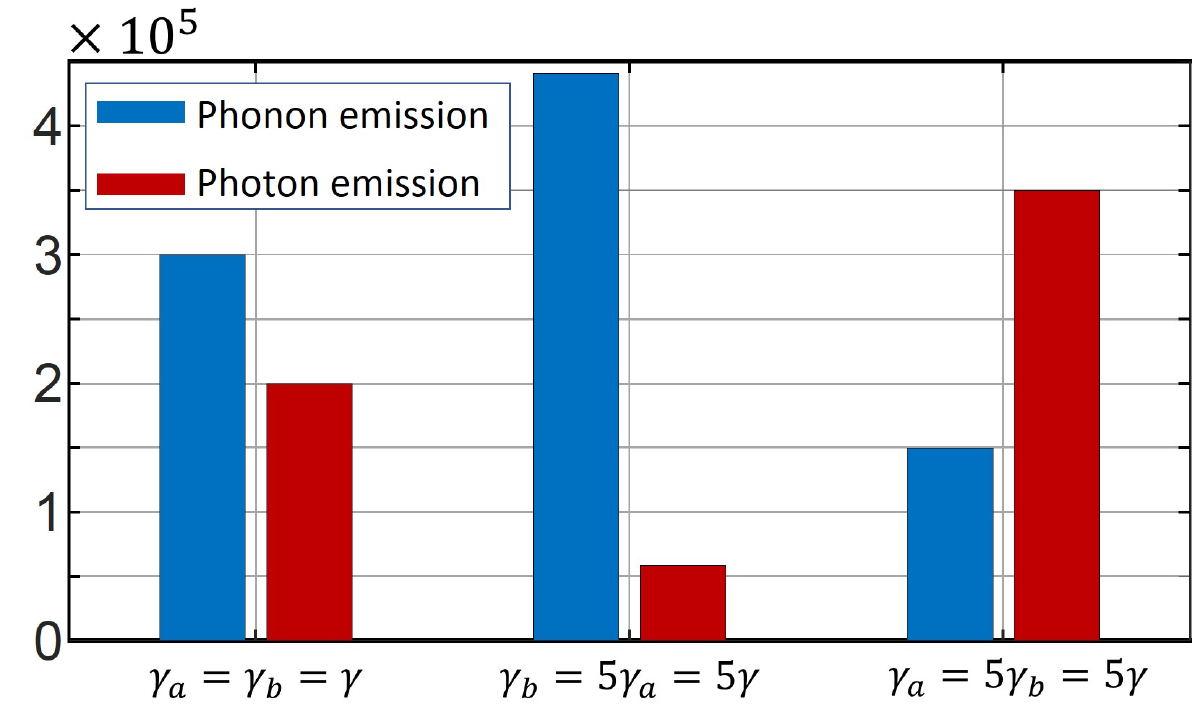}
		\caption{Statistical results of photon emission trajectory number and phonon emission trajectory number under different dissipation rates. The dissipation rate $\gamma=10^{-9}\omega_{m}$ and other parameters are chosen as in Fig.~\ref{F4}.}
		\label{F6}
	\end{figure}
    
	By using the simulation of quantum trajectories, it can be found that a photon (phonon) jumping probability  $P_{a} (P_{b})$ is related to the photon (phonon) excitation number. As illustrated in Fig.~\ref{F5}, when the average excitation number is greater than 1, the corresponding photon (or phonon) emission probability has a relatively significant value.  As the number of phonons (or photons) within the system decreases, the probability of the system emitting the corresponding particles also decreases. This result is consistent with physical reality~\cite{Marlan1997,walls2008optics}. If calculations are performed using the master equation, such an intuitive simulation result cannot be obtained. {Leveraging  the characteristic that the probability of photon (phonon) jumping changes with the photon (phonon) excitation number, one can reconstruct the three-phonon-two-photon oscillation process~\cite{PhysRevA.105.023720}. Such a reconstruction of the process is shown in Fig.~\ref{F12}. The characteristics of the energy exchange can be determined with arbitrary precision by collecting sufficient data on quantum jumps from the cavity and mechanical oscillator.}

     In the quantum trajectory simulation, the selection of the system's dissipation channel is found to be proportional to the dissipation rate of the corresponding mode. This implies that as the photon (phonon) dissipation rate increases, the number of trajectories that emit photons (phonons) also increases. This is exemplified by the statistical data presented in Fig.~\ref{F6}. When the phonon dissipation rate is five times larger than the photon dissipation rate, the number of photon emission trajectories is markedly smaller than the number of phonon emission trajectories. When the phonon dissipation rate is equal to or less than the photon dissipation rate, the number of photon emission trajectories becomes significant. This suggests that increasing the photon dissipation rate appropriately can enhance the number of photon emission trajectories.

	\subsection{$N$-photon/$n$-phonon emission }

\begin{figure}
	\centering
	\includegraphics[scale=0.6]{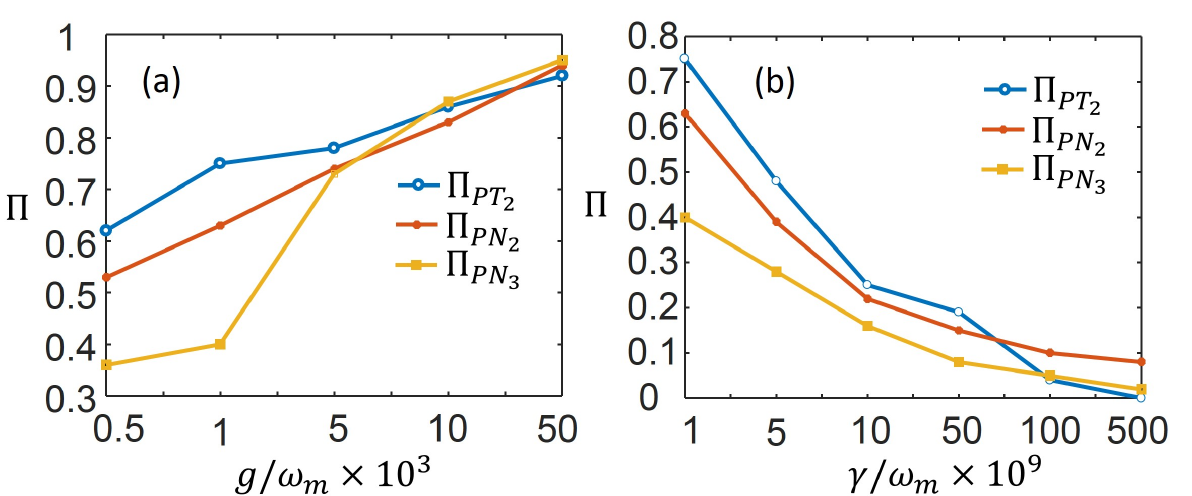}
	\caption{Purities of two-photon, two-phonon, and three-phonon emissions vs (a) g/$\omega_{m}$ and (b) $\gamma/\omega_{m}$. When the coupling strength exceeds $0.005\omega_{m}$, the three-photon-two-photon resonance conditions that the system needs to satisfy are obtained through numerical calculations. In panel (a), system parameters are $\gamma_{a}=\gamma_{b}=10^{-9}\omega_{m}$. In panel (b), system parameters are $g=0.001\omega_{m}$ and $\omega_{c}=1.500105\omega_{m}$.}
	\label{F7}
\end{figure}
    
	An intriguing physical phenomenon depicted in Fig.~\ref{F4} warrants further investigation. When the time interval between the emission of two photons (two phonons) is shorter than the photon lifetime $\tau_{a}=1/\gamma_{a}$ (or the phonon lifetime $\tau_{b}=1/\gamma_{b}$), such an emission can be considered a two-photon (two-phonon) emission~\cite{PhysRevLett.124.053601,PhysRevLett.133.043601}. In particular, when the total time required for the emission of three phonons is less than the lifetime of a photon, this indicates the occurrence of three-phonon emission. We define the purities of $n$-photon emission $\Pi_{PT_{n}}$ and $n$-phonon emission $\Pi_{PN_{n}}$ as
    \begin{equation}
        \Pi_{PT_{n}}=\frac{N_{PT_{n}}}{N_{PT}},\ \ \Pi_{PN_{n}}=\frac{N_{PN_{n}}}{N_{PN}},
    \end{equation}
respectively, where $N_{PT_{n}} (N_{PN_{n}})$ is the number of trajectories that produces $n$-photon ($n$-phonon) emission, $N_{PT} (N_{PN})$ is the number of trajectories that emit photons (phonons).

In Fig.~\ref{F7}, it is evident that the purities of the two-photon, two-phonon, and three-phonon emissions are significantly affected by both the coupling strength and the dissipation rate. When the coupling strength reaches $g/\omega_{m}=0.05$ and the dissipation rate $\gamma_{a}=\gamma_{b}=10^{-9}\omega_{m}$,  the purities of the two-photon, two-phonon, and three-phonon emissions are 0.92, 0.94, and 0.95, respectively. {The underlying mechanism for this phenomenon can be explained as follows: (i) An increase in the phonon dissipation rate $\gamma_{b}$ or photon dissipation rate $\gamma_{a}$ directly corresponds to an acceleration in the rate at which phonons or photons escape from the system. This leads to a corresponding shortening of their average lifetimes $\tau_{b}$ or $\tau_{a}$, thereby increasing the probability of bundle emission in a short period of time~\cite{Muoz2014,PhysRevLett.124.053601,PhysRevLett.133.043601}. (ii) An increase in the coupling strength $g$ can enhance the effective coupling strength between the states $\ket{0,3}$ and $\ket{2,0}$. This decreases the average time interval between photon (phonon) emission events, further boosting the probability of phonon and photon bundle emissions~\cite{PhysRevLett.124.053601,PhysRevLett.133.043601}.} Compared to the results obtained from the master equation simulations~\cite{PhysRevX.8.011031}, the quantum trajectory method provides a deeper insight into the emission mechanisms of phonons and photons once the dissipative channel is opened (e.g., two-photon emission, two-phonon emission, and three-phonon emission). Unlike the master equation approach, which primarily characterizes the correlations between emitted particles using correlation functions, the quantum trajectory method enables a more detailed understanding of the underlying causes of particle correlations at a microscopic level.

\begin{figure}
    \centering
    \includegraphics[scale=0.41]{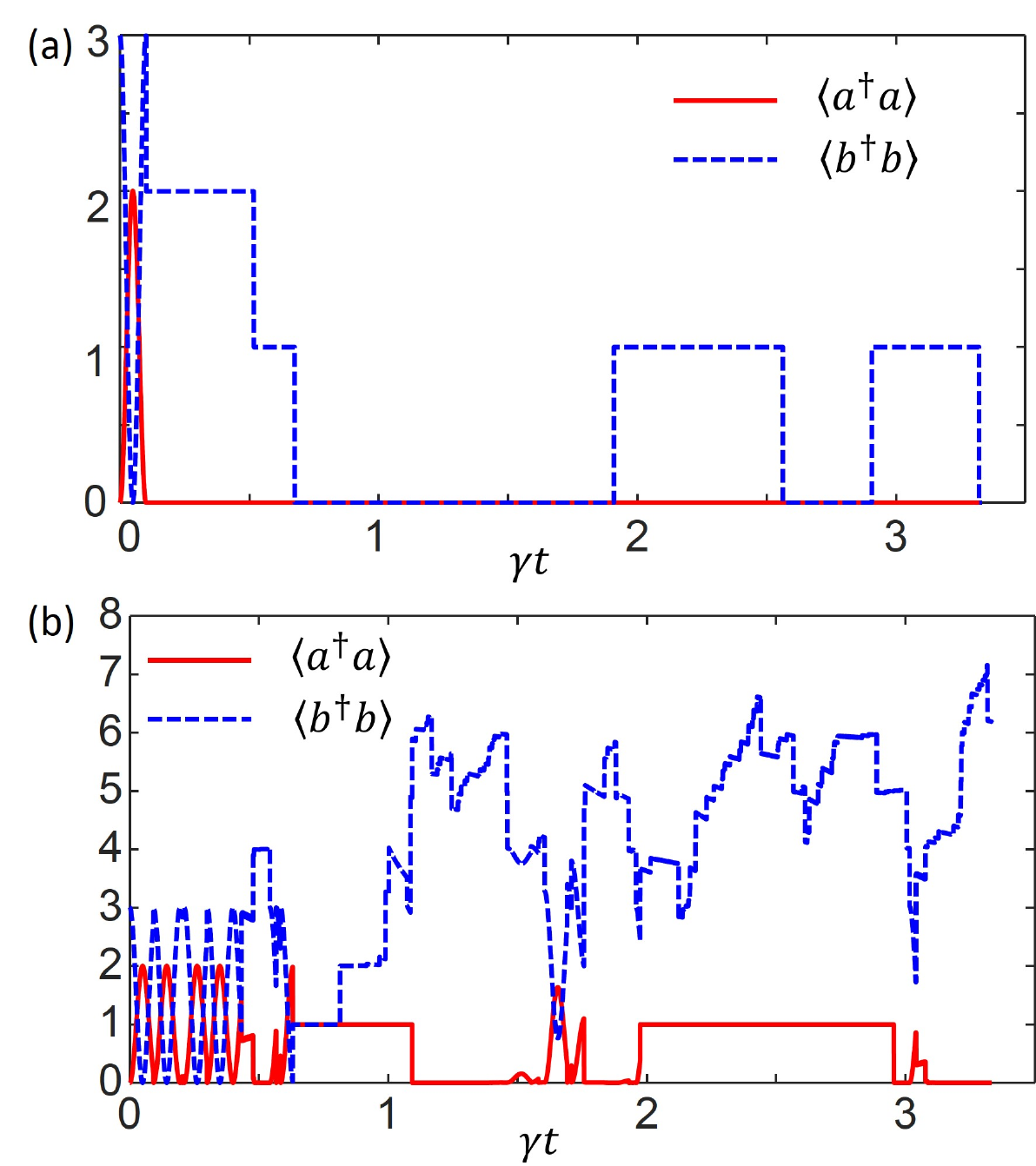}
    \caption{Examples of a single quantum trajectory. In panel (a), the temperature of the environment is $T=\hbar \omega_{m}/k$. In panel (b), the temperature of the environment is $T=3\hbar \omega_{m}/k$. In both panels, other parameters are $\gamma=\gamma_{a}=\gamma_{b}=3\times10^{-9}\omega_{m}$, $g=0.001\omega_{m}$, and $\omega_{c}=1.5000105\omega_{m}$. }
    \label{99}
\end{figure}
\begin{figure}
    \centering
    \includegraphics[scale=0.55]{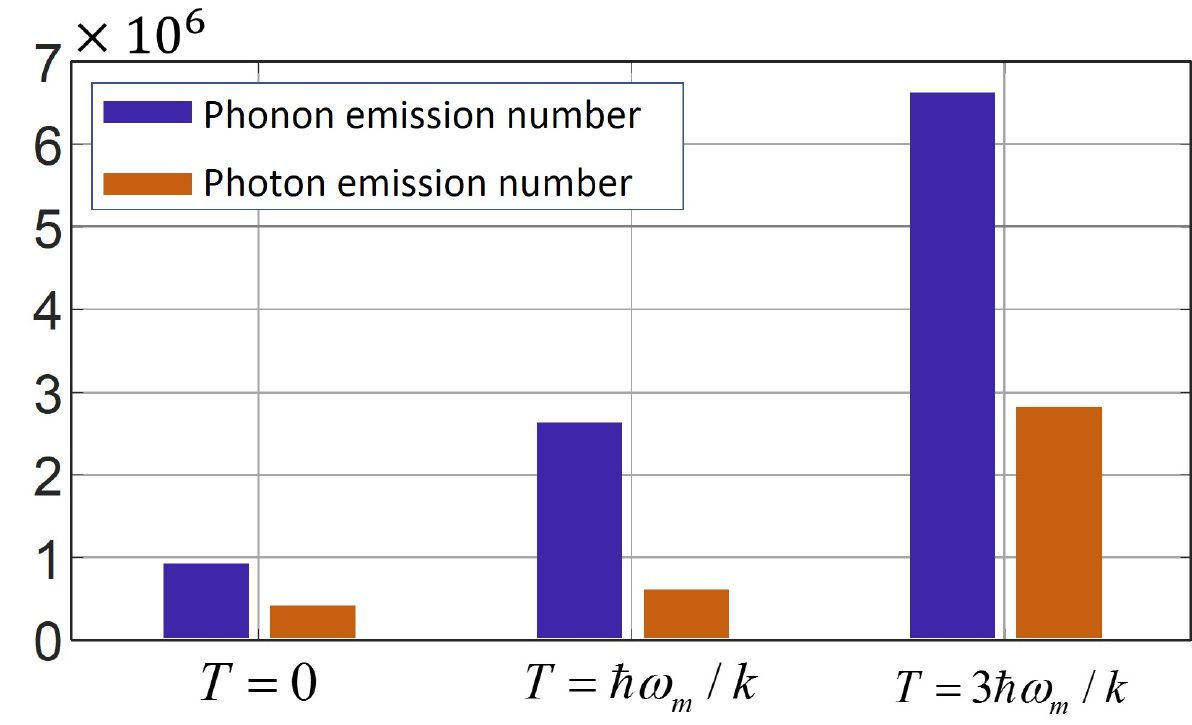}
    \caption{Statistical results of photon emission number and phonon emission number under different temperature. System parameters are $\gamma=\gamma_{a}=\gamma_{b}= 10^{-9}\omega_{m}, g=0.001\omega_{m}$, and $\omega_{c}=1.5000105\omega_{m}$. We perform $5\times10^{5}$ quantum trajectory simulations at each temperature, with each trajectory simulation time range limited to between 0 and $3/\gamma$. The symbols $k$ and $\hbar$ represent the Boltzmann constant and Planck constant, respectively. When $\omega_{m}/2\pi \sim1$GHz, $\hbar\omega_{m}/k \sim 0.5K$, where $K$ denotes the temperature unit in Kelvin. }
    \label{F101}
\end{figure}
\begin{figure}
    \centering
    \includegraphics[scale=0.55]{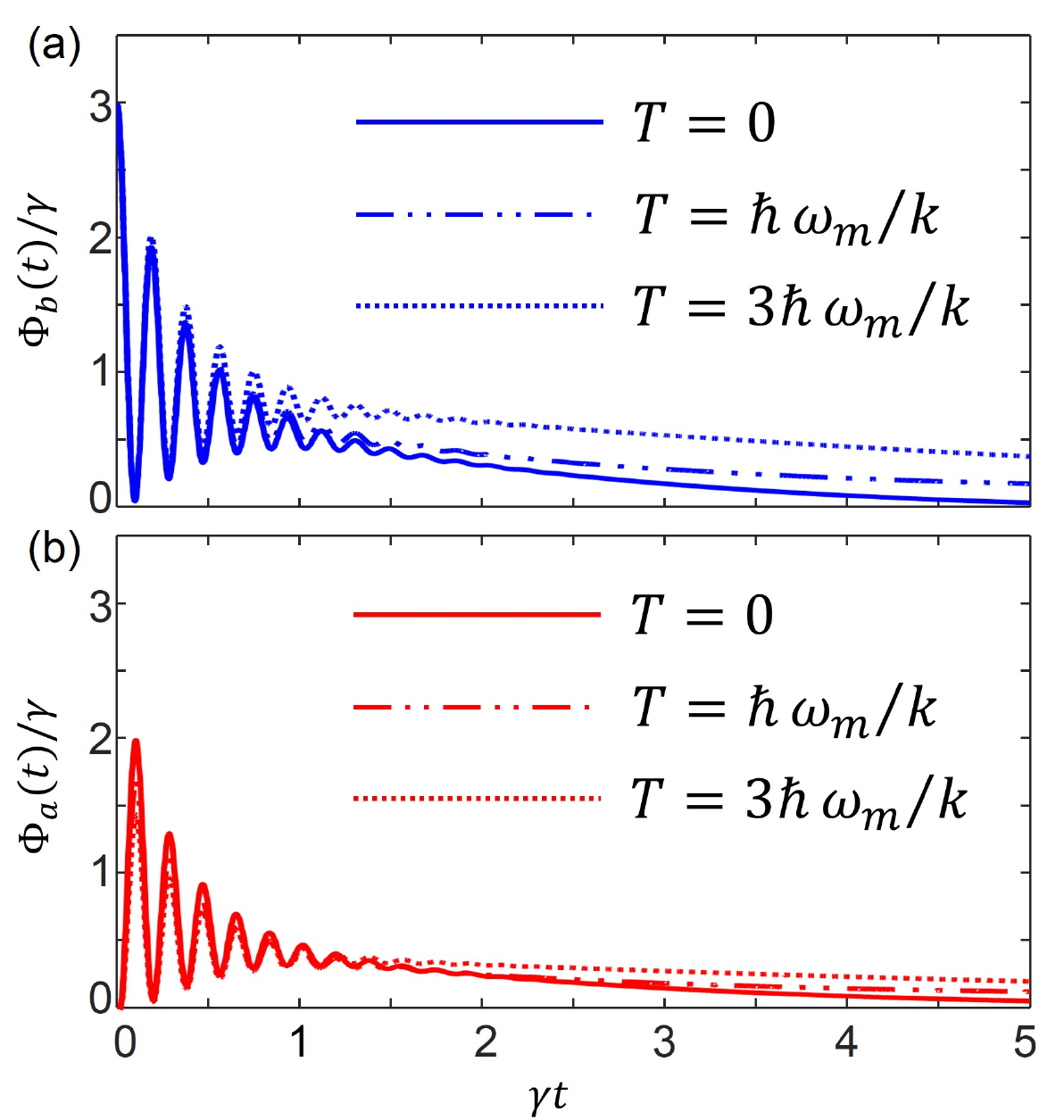}
    \caption{Plots of the output photon flux $\Phi_{a}(t)$ and the output phonon flux $\Phi_{b}(t)$ as functions of time, for different temperature. System parameters are $\gamma=\gamma_{a}=\gamma_{b}= 10^{-9}\omega_{m}, g=0.001\omega_{m}$, and $\omega_{c}=1.5000105\omega_{m}$.}
    \label{F102}
\end{figure}

   \subsection{The effect of temperature on spontaneous emission}
    Typically, the quantum systems we study are considered in a zero temperature environment, where the master equation accounts for single-photon (single-phonon) loss and the corresponding dephasing effects induced by the environment~\cite{Marlan1997,walls2008optics}. However, in actual experiments, the environment always has a finite temperature, which introduces discrepancies between experimental results and theoretical predictions. In this subsection, we examine the effects of a finite-temperature environment on photon and phonon emission.

    For simplicity, we focus solely on the effect of temperature on mechanical mode. Therefore, in the quantum trajectory simulation, the system's jumping operators are expressed by these three terms~\cite{PhysRevA.46.4382}:
    \begin{eqnarray}
    \nonumber
        L_{1}&=&\sqrt{\gamma_{b}n(\omega_{m})+\gamma_{b}}b,\\ \nonumber
        L_{2}&=&\sqrt{\gamma_{b}n(\omega_{m})}b^{\dagger},\\ 
        L_{3}&=&\sqrt{\gamma_{a}}a,
        \label{e13}
    \end{eqnarray}
    where
    \begin{equation}
    n(\omega_{m}) = \frac{1}{e^{\frac{\hbar \omega_{m}}{kT}} - 1}
    \label{n12}
    \end{equation}
    is the Bose-Einstein distribution function, representing the average number of thermal phonons in the environment at frequency $\omega_{m}$ and temperature $T$. In Eq.~(\ref{n12}), $k$ and $\hbar$ represent the Boltzmann constant and the Planck constant, respectively. 
    
    Figure~\ref{99} shows two examples of a single trajectory at different temperatures. At lower temperatures [see in Fig.~\ref{99}(a)], thermal noise in the environment excites phonons in the system. However, the excited phonon number is insufficient to initiate the three-phonon-two-photon oscillation process, resulting in a negligible effect on the number of photons emitted. As the temperature increases [see in Fig.~\ref{99}(b)], the number of phonons excited in the system rises significantly, which in turn greatly enhances the probability of photon emission. Statistical analysis on a large number of trajectories is shown in Fig.~\ref{F101}. {It is evident that as the temperature increases, the number of emitted photons and phonons over the entire simulation time also rises. In Fig.~\ref{F102}, we illustrate the simulation result of the photon flux $\Phi_{a}(t)$ and the phonon flux $\Phi_{b}(t)$. The system output flux are defined by
    
    \begin{equation}
    \Phi_{a(b)}(t) = \gamma_{a(b)} \text{Tr}[\rho(t) X^{\dagger}_{a(b)} X_{a(b)}],
\end{equation}
where $X_{a}=a$, $X_{b}=b$, and $\rho(t)$ is the density matrix of the system. It can be observed that an increase in temperature results in a higher output flux of both photons and phonons in the system, which aligns with the statistical result derived using the quantum trajectory approach, as shown in Fig.~\ref{F101}.}
    
    It should be noted that we only considered thermal noise of a specific frequency in the environment.  In this context, the environment injects energy into the system with a certain probability. This leads to an increase the number of photon and phonon emissions when the temperature rises (i.e., when the probability of energy input into the system increases). However, an increase in the number of photons and phonons does not necessarily imply an increase in the purity of two-photon emission, two-phonon emission, and three-phonon emission. As can be seen from Eq.~(\ref{e13}), when the temperature increases, the corresponding phonon dissipation rate also increases, which clearly reduces the purity of n-phonon/n-photon emission. The simulation results indicate that when the temperature is $\hbar \omega
    _{m}/k$ ($g=0.05\omega_{m}$, $\gamma_{a}=\gamma_{b}= 10^{-9}\omega_{m}$), the purities of two-photon emission, two-phonon emission, and three-phonon emission are 0.91, 0.91, and 0.93, respectively. When the temperature is $3\hbar \omega
    _{m}/k$ ($g=0.05\omega_{m}$, $\gamma_{a}=\gamma_{b}= 10^{-9}\omega_{m}$), the corresponding purities decrease to 0.71, 0.61, and 0.38, respectively. In typical optomechanical experiments, the ambient temperature is much lower than $\hbar \omega_{m}/k$ (for $\omega_{m}/2\pi \sim1$GHz, $\hbar\omega_{m}/k \sim 0.5K$, where $K$ denotes the temperature unit in Kelvin)~\cite{PhysRevA.90.011803,PhysRevX.5.041002,Riedinger2018}, well below the threshold required to induce significant thermal effects, thereby enabling the observation of $n$-photon/$n$-phonon emission phenomena without the need for stringent low-temperature conditions. 
    
\subsection{Parametric sensitivity} 
\begin{figure}
		\centering
		\includegraphics[scale=0.5]{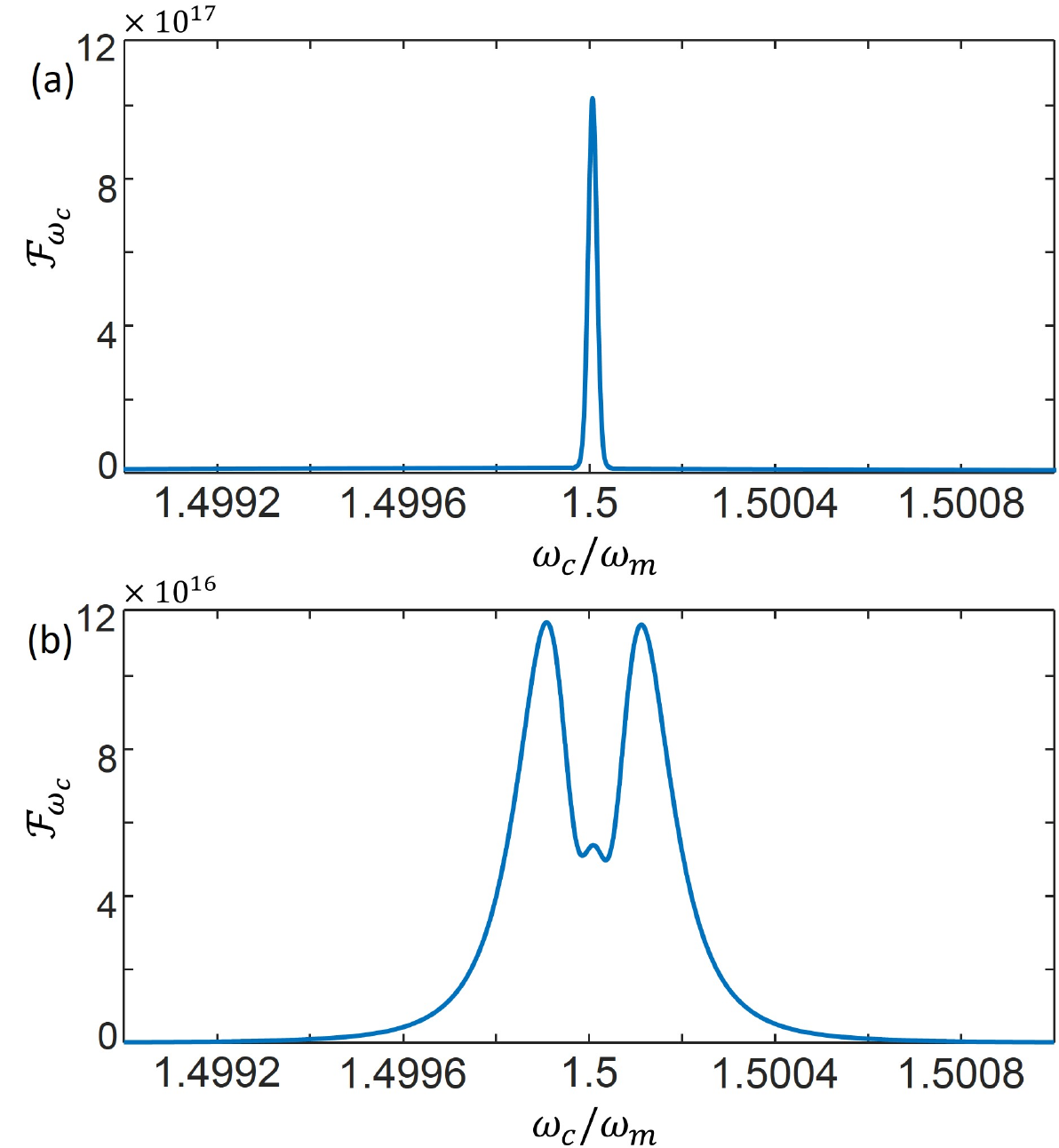}
		\caption{Quantum Fisher Information (QFI)  $\mathcal{F}_{\omega_{c}}$ as a function of $\omega_{c}$ for  (a) the total Hamiltonian $H_{s}$ and (b) the effective Hamiltonian $H_{\rm eff}$ .  Other parameters are $g=0.001\omega_{m}$ and $t_{f}=1.0077\times10^{8}/\omega_{m}$.}
		\label{F9}
	\end{figure}
	The resonance between the three-phonon state  and the two-photon state is sensitive to the parameters, which can be reflected in the size of the quantum Fischer information (QFI)~\cite{PhysRevLett.124.120504,PhysRevA.106.062616}. Since the system is in a pure state, the QFI relative to
	the parameter of interest $A$ can be calculated as~\cite{PhysRevLett.124.120504} 
	\begin{equation}
		\mathcal{F}_{A}=4\left(\left\langle\partial_{A} \phi \mid \partial_{A} \phi\right\rangle-\left|\left\langle\phi \mid \partial_{A} \phi\right\rangle\right|^{2}\right),
	\end{equation}
	where $\ket{\phi}$ is the state to be measured. We are interested in the sensitivity of the system evolution to the frequency of the cavity field, so the parameter $A$ is chosen as $\omega_{c}$ and the state
	\begin{equation}
		\ket{\phi}={\rm exp}^{-iH t_{f}}\ket{0,3}.
	\end{equation}
	In Fig.~\ref{F9}, we plot QFI as  a function of $\omega_{c}$ for  the total Hamiltonian (i.e. $H=H_{s}$) in panel (a) and the effective Hamiltonian (i.e. $H=H_{\rm eff}$) in panel (b). It can be observed that the dynamics governed by the effective Hamiltonian $H_{\rm eff}$ exhibits less sensitivity to the parameter $\omega_{c}$ compared to those governed by the total Hamiltonian $H_{s}$. This difference arises because many fast oscillation terms, which are sensitive to parameters, are neglected in the process of deriving the effective Hamiltonian. Note that QFI for the total Hamiltonian remains above  $10^{16}$, indicating that the three-photon-two-photon evolution process is a high-order phenomenon with a strong sensitivity to parameters. This somewhat hinders the experimental observation of photon emission induced by mechanically excited states. Nevertheless, the aforementioned sensitivity offers the potential for precision measurements~\cite{PhysRevLett.96.010401,PhysRevLett.124.120504} to be conducted in an experimental setting.

	\section{EXPERIMENTAL IMPLEMENTATION} \label{s4}
	
	As depicted in Fig.~\ref{F7}, a superconducting quantum interference device (SQUID) loop comprising a mechanical vibrating arm is embedded into a coplanar microwave cavity of length $d$ and frequency $\omega_{c1}$, thereby enabling the effective coupling between the mechanical oscillator and the cavity~\cite{Niemczyk2010,RevModPhys.91.025005,FriskKockum2019,RevModPhys.93.025005,PhysRevX.2.021007,PhysRevA.93.022510}. The SQUID  can be modeled as a lumped circuit element at the center $x=0$ of the cavity. The mechanical oscillator can be modeled as a harmonic oscillator, with the $y$ coordinate representing the centre-of-mass displacement, an oscillation frequency of $\omega_{m1}$, and a mass of $m$. The magnetic flux threading the SQUID loop is given by $\Phi_{\rm ext}(y)=\Phi_{\rm ext}(0)+\lambda B_{\rm osc}l_{\rm osc}y$, where $\Phi_{\rm ext}(0) \equiv \Phi_{\rm ext0}$ is the flux with the mechanical oscillator fixed at $y=0$, $B_{\rm osc}$ is the local magnetic field in the vicinity of the mechanical resonator, $\lambda$ is a dimensionless geometrical factor that accounts for the nonuniform displacement of the resonator along its extension. The Hamiltonian, describing the dynamics for the cavity, SQUID, and mechanical oscillator is given by ($\hbar=1$)
	\begin{eqnarray}
		\nonumber
		H_{cSm}&=&\omega_{c1}^{\rm dc}a^{\dagger}a+\omega_{m1}b^{\dagger}b
		-\frac{\alpha}{2}(e^{i\omega_{d}t}+e^{-i\omega_{d}t})(a+a^{\dagger})^2\\
		&&-\frac{g_{0}}{2}(a+a^{\dagger})^2(b+b^{\dagger}),
		\end{eqnarray}
    where the effective cavity frequency is defined to be $\omega_{c1}^{\rm dc}\equiv \sqrt{4/C_{c}dL_{J}(\Phi_{\rm dc})}$. The coupling strengths are expressed as 
    \begin{equation}
    	\alpha \equiv \frac{\omega_{c1}^{\rm dc}}{4}\frac{\pi \delta\Phi}{\Phi_{0}}\tan\left(\frac{\pi \Phi_{\rm dc}}{\Phi_{0}}\right)
    \end{equation}
    and
    \begin{equation}
    	g_{0}\equiv \omega_{c1}^{\rm dc}\frac{\lambda B_{\rm osc}l_{\rm osc}y_{\rm zp}}{\Phi_{0}/\pi}\frac{L_{J}(\Phi_{\rm dc})}{L_{c}d}\tan\left(\frac{\pi\Phi_{\rm dc}}{\Phi_{0}}\right),
    \end{equation}
    where $\Phi_{0}=h/2e$ is the flux quantum and $y_{\rm zp}= \sqrt{1/2m\omega_{m1}}$ is the zero-point displacement of the mechanical resonator. Here we suppose that the external flux is weakly modulated around some fixed dc bias $\Phi_{\rm ext0}=\Phi_{\rm dc}+\delta\Phi\cos(\omega_{d}t)$. The specific quantisation processes of the circuit in Fig.~\ref{F7} can be found in Ref.~\cite{PhysRevA.93.022510}. When the effective cavity frequency and the mechanical oscillation frequency satisfy the resonance condition in Appendix~\ref{app:a}, we can obtain the same third-order effective Hamiltonian as in Eq.~(\ref{eq3}). 
\begin{figure}
		\centering
		\includegraphics[scale=0.4]{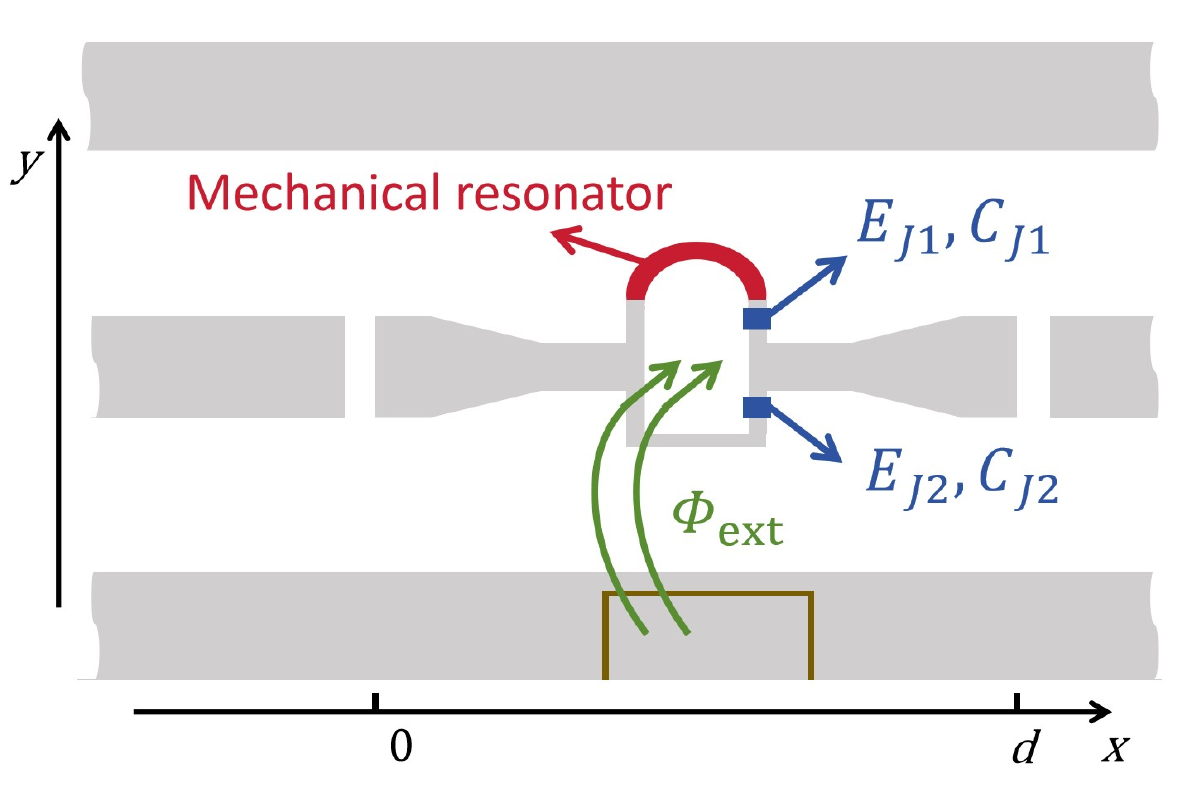}
		\caption{Layout of a coplanar microwave cavity, characterized by its inductance and capacitance per unit length $L_{c}$ and $C_{c}$, respectively, bisected by a dc SQUID. The SQUID is driven by a time-dependent external flux $\Phi_{\rm ext}(t)$ (green) and constituted of two Josephson junctions (blue), which are distinguished by their energy $E_{J}$, and capacitance $C_{J}$. One arm of the SQUID loop is mechanically compliant, forming a doubly clamped resonator (red) of length $l_{\rm res}$. For purposes of clarity, the SQUID has been greatly enlarged in this representation with respect to the superconducting cavity.}
		\label{F8}
	\end{figure}
    
	In the preceding sections, we have assumed that the mechanical oscillator is in an excited state (mechanical Fock state). This state is possible to be prepared, for example, the mechanical oscillator in the SQUID is coupled to an additional qubit~\cite{OConnell2010}, using the same protocols realized in circuit QED~\cite{Hofheinz2009}. In the above simulations, we have chosen very low optical cavity dissipation rates and mechanical oscillator dissipation rates because the three-phonon-to-two-photon transition is a high-order nonlinear process with a weak effective coupling strength. To observe two-photon emission as well as two-phonon (and three-phonon) emission, very long photon lifetimes in the cavity and phonon lifetimes in the mechanical oscillator are required. State-of-the-art 3D cavities currently exhibit photon lifetimes on the order of milliseconds~\cite{PRXQuantum.5.040307}, while the quality factor of nanomechanical oscillators exceeds $10^{10}$~\cite{PhysRevLett.132.133602}. Therefore, the simulation parameters we have selected are experimentally feasible.

	\section{CONCLUSIONS} \label{s6}
	We theoretically derive the effective Hamiltonian for three-phonon-two-photon oscillations under the weak coupling condition, enabling accurate calculation of the resonance frequency and effective coupling strength. Using the quantum trajectory method, we explore the phonon and photon emission mechanisms in this higher-order process within an open system. Simulation results show that when the cavity field or mechanical oscillator is in a highly excited state, the corresponding emission probability reaches its maximum. Furthermore, we investicate the purity of two-photon, two-phonon, and three-phonon emissions throughout the emission process. The results indicate that the purity of these emissions increases with stronger coupling strength and lower dissipation rates. We also discuss the influence of environmental thermal noise on the emission mechanism, with simulation results revealing that as the temperature increases, both the number of photon and phonon emissions rise. Finally, we propose an experimental setup that could enable the implementation of this protocol. This study provides deeper insights into the mechanism between phonon emission and photon emission in the DCE.
	
	 \section*{Acknowledgements}
	We express sincere gratitude to Qian Bin and Yang Liu for the insightful discussions and valuable suggestions. Yan Xia was supported by the National Natural Science Foundation of China under Grant No. 62471143, the Key Program of National Natural Science Foundation of Fujian Province under Grant No. 2024J02008, and the project from Fuzhou University under Grant No. JG2020001-2. Y.-H.C. was supported by the National Natural Science Foundation of China under Grant No. 12304390.

    \section*{Disclosures}
    The authors declare no conflicts of interest.

    \section*{Data availability}
     Data underlying the results presented in this paper are not publicly available at this time but maybe obtained from the authors upon reasonable request.

	\appendix
    	\section{The derivation of the effective Hamiltonian} \label{app:a}
    	To apply the generalized Jame's effective Hamiltonian method~\cite{PhysRevA.95.032124} to Eq.~(\ref{eq1}) in the main text, we first rewrite it in the interaction picture,
    	\begin{eqnarray}
    		\nonumber
    		H_{I}(t)&=&g[a^{\dagger2}b^{\dagger}e^{i(2\omega_{c}+\omega_{m})t}+a^{\dagger2}be^{i(2\omega_{c}-\omega_{m})t}
    		+(2a^{\dagger}a+1)b^{\dagger}e^{i\omega_{m}t}]+{\rm H.c.}\\
    	\end{eqnarray}
    	This can also be written in the following form:
    	\begin{eqnarray}
    		H_{I}(t)=\sum_{k=1}^{3}[h_{k}e^{i\omega_{k}t}+{\rm H.c.}],
    	\end{eqnarray}
    	where
    	\begin{eqnarray}
    		\begin{array}{ll}
    			h_{1}=g(a^{\dagger})^2b^{\dagger}, & \omega_{1}=2\omega_{c}+\omega_{m},\\
    			h_{2}=g(a^{\dagger})^2b, & \omega_{2}=2\omega_{c}-\omega_{m},\\
    			h_{3}=g(2a^{\dagger}a+1)b^{\dagger}, & \omega_{3}=\omega_{m}.
    		\end{array}
    	\end{eqnarray}
    	In case of all the frequencies $\omega_{k}$ being distinct , the second-order and the third-order effective Hamiltonian can be obtained by calculating the following formulas:
    	\begin{eqnarray}
    		H_{\rm eff}^{(2)}=\sum_{k}\frac{1}{\omega_{k}}[h_{k},h_{k}^{\dagger}],
    	\end{eqnarray}
        \begin{equation}
        	\begin{aligned}
        		H_{\rm eff}^{(3)}= &  \sum_{l, k, n}\left\{\frac{1}{\omega_{n}\left(\omega_{n}-\omega_{k}\right)}\left[h_{l} h_{k}^{\dagger} h_{n} e^{i\left(\omega_{l}-\omega_{k}+\omega_{n}\right) t}+h_{l} h_{k} h_{n}^{\dagger} e^{i\left(\omega_{l}+\omega_{k}-\omega_{n}\right) t}+{\rm H.c.}\right]\right. \\
        		& \left.+\frac{1}{\omega_{n}\left(\omega_{n}+\omega_{k}\right)}\left[h_{l}^{\dagger} h_{k} h_{n} e^{i\left(-\omega_{l}+\omega_{k}+\omega_{n}\right) t}+{\rm H.c.}\right]\right\}.
        	\end{aligned}
        \end{equation}
        Considering the resonant condition $2\omega_{c}=3\omega_{m}$, all frequency contributions which are different from zero can be neglected by performing the rotating-wave approximation, so we have
        \begin{eqnarray}
        	\nonumber
        	H_{\rm eff}^{(2)}&=&\frac{1}{\omega_{1}}[h_{1},h_{1}^{\dagger}]+\frac{1}{\omega_{2}}[h_{2},h_{2}^{\dagger}]+\frac{1}{\omega_{3}}[h_{3},h_{3}^{\dagger}],\\
        \end{eqnarray}
        \begin{eqnarray}
        	\nonumber
        	H_{\rm eff}^{(3)}&=&\frac{1}{\omega_{2}(\omega_{2}-\omega_{1})}[h_{2}h_{1}^{\dagger}h_{2}+h_{2}^{\dagger}h_{1}h_{2}^{\dagger}]+\frac{1}{\omega_{1}(\omega_{1}-\omega_{2})}[h_{2}h_{2}h_{1}^{\dagger}+h_{2}^{\dagger}h_{2}^{\dagger}h_{1}]\\\nonumber&&+\frac{1}{\omega_{2}(\omega_{2}+\omega_{2})}[h_{1}^{\dagger}h_{2}h_{2}+h_{1}h_{2}^{\dagger}h_{2}^{\dagger}]
        	+\frac{1}{\omega_{3}(\omega_{3}-\omega_{2})}[h_{3}h_{2}^{\dagger}h_{3}+h_{3}^{\dagger}h_{2}h_{3}^{\dagger}]\\\nonumber&&+\frac{2}{\omega_{2}(\omega_{2}-\omega_{3})}[h_{3}h_{3}h_{2}^{\dagger}+h_{3}^{\dagger}h_{3}^{\dagger}h_{2}]+\frac{1}{\omega_{3}(\omega_{3}+\omega_{3})}[h_{2}^{\dagger}h_{3}h_{3}+h_{2}h_{3}^{\dagger}h_{3}^{\dagger}],\\
        \end{eqnarray}
        where the term $h_{2}h_{1}^{\dagger}h_{2}$ represents the virtual transition path $\ket{n,k}\rightarrow\ket{n+2,k-1}\rightarrow\ket{n,k-2}\rightarrow\ket{n+2,k-3}$, the term $h_{2}h_{2}h_{1}^{\dagger}$ represents the virtual transition path $\ket{n,k}\rightarrow\ket{n+2,k-1}\rightarrow\ket{n+4,k-2}\rightarrow\ket{n+2,k-3}$, the term $h_{1}^{\dagger}h_{2}h_{2}$ represents the virtual transition path $\ket{n,k}\rightarrow\ket{n-2,k-1}\rightarrow\ket{n,k-2}\rightarrow\ket{n+2,k-3}$, the term $h_{3}^{\dagger}h_{2}h_{3}^{\dagger}$ represents the virtual transition path $\ket{n,k}\rightarrow\ket{n,k-1}\rightarrow\ket{n+2,k-2}\rightarrow\ket{n+2,k-3}$, the term $h_{3}^{\dagger}h_{3}^{\dagger}h_{2}$ represents the virtual transition path $\ket{n,k}\rightarrow\ket{n,k-1}\rightarrow\ket{n,k-2}\rightarrow\ket{n+2,k-3}$, the term $h_{2}h_{3}^{\dagger}h_{3}^{\dagger}$ represents the virtual transition path $\ket{n,k}\rightarrow\ket{n+2,k-1}\rightarrow\ket{n+2,k-2}\rightarrow\ket{n+2,k-3}$ (where $n$ and $k$ are the number of photon excitations and phonon excitations, respectively).
        Finally, we get James' effective Hamiltonian as follow
        \begin{eqnarray}
        \label{21}
        	H_{\rm eff}&=&\omega_{c}a^{\dagger}a+\omega_{m}b^{\dagger}b+\frac{g^{2}}{4\omega_{m}}\left[a^{\dagger2}a^{2}-2(2a^{\dagger}a+1)(3b^{\dagger}b+4a^{\dagger}a+3)\right]\\\nonumber
            &&+\frac{9g^{3}}{\omega_{m}^{2}}(a^{\dagger2}b^{3}+a^{2}b^{\dagger3})
        \end{eqnarray}

   Note that adiabatic elimination~\cite{PhysRevA.92.023842} and Schrieffer-Wolff transformation~\cite{PhysRev.149.491} are also powerful tools to derive the effective Hamiltonian. Here, we briefly review these two computational frameworks. 
  
  (1) For the approach of adiabatic elimination, one should reduce the system Hamiltonian in the truncated Hilbert space spanned by the bare states $\ket{0,1}$, $\ket{0,2}$, $\ket{0,3}$, $\ket{2,0}$, $\ket{2,1}$, $\ket{2,2}$, and $\ket{4,1}$. Moving to the rotating frame with frequency $2\omega_{c}$, one can obtain the transformed reduced Hamiltonian

  \begin{equation}
      H_{r} = 
\begin{pmatrix}
\omega_m-2\omega_c & \sqrt{2}g & 0 & \sqrt{2}g & 0 & 2g & 0 \\
\sqrt{2}g & 2\omega_m-2\omega_c & \sqrt{3}g & 0 & 2g & 0 & 0 \\
0 & \sqrt{3}g & 3\omega_m-2\omega_c & 0 & 0 & \sqrt{6}g & 0 \\
\sqrt{2}g & 0 & 0 & 0 & 5g & 0 & 2\sqrt{3}g \\
0 & 2g & 0 & 5g &  \omega_m & 5\sqrt{2}g & 0 \\
2g & 0 & \sqrt{6}g & 0 & 5\sqrt{2}g &  2\omega_m & 2\sqrt{6}g \\
0 & 0 & 0 & 2\sqrt{3}g & 0 & 2\sqrt{6}g & 2\omega_c + \omega_m
\end{pmatrix}
  \end{equation}
After the transformation, an arbitrary state of the system in this truncated Hilbert space can be denoted by $(c1,c2,c3,c4,c5,c6,c7)^{T}$ and the Schr\"{o}dinger equation with Hamiltonian $H_{r}$ gives
\begin{align}
\label{0}
i\dot{c}_{1} &= (\omega_{m} - 2\omega_{c}) c_{1} + \sqrt{2}g\, c_{2} + \sqrt{2}g\, c_{4} + 2g\, c_{6}, \\ \label{1}
i\dot{c}_{2} &= \sqrt{2}g\, c_{1} + (2\omega_{m} - 2\omega_{c}) c_{2} + \sqrt{3}g\, c_{3} + 2g\, c_{5}, \\ \label{2}
i\dot{c}_{3} &= \sqrt{3}g\, c_{2} + (3\omega_{m} - 2\omega_{c}) c_{3} + \sqrt{6}g\, c_{6}, \\ \label{3}
i\dot{c}_{4} &= \sqrt{2}g\, c_{1} + 5g\, c_{5} + 2\sqrt{3}g\, c_{7}, \\ \label{4}
i\dot{c}_{5} &= 2g\, c_{2} + 5g\, c_{4} + \omega_{m} c_{5} + 5\sqrt{2}g\, c_{6}, \\ \label{5}
i\dot{c}_{6} &= 2g\, c_{1} + \sqrt{6}g\, c_{3} + 5\sqrt{2}g\, c_{5} + 2\omega_{m} c_{6} + 2\sqrt{6}g\, c_{7}, \\ \label{6}
i\dot{c}_{7} &= 2\sqrt{3}g\, c_{4} + 2\sqrt{6}g\, c_{6} + \left(2\omega_{c} + \omega_{m}\right) c_{7}. 
\end{align}
For the condition $g/\omega_{m}\ll1$, the adiabatic elimination in  Eqs.~(\ref{0})-(\ref{1}) and (\ref{4})-(\ref{6}) can be applied and the coefficients $c1$, $c2$, $c5$, $c6$ and $c7$ can be obtained by . Substituting these results in Eqs.~(\ref{2}) and (\ref{3}) and considering the near-resonance case $2\omega_{c}\approx3\omega_{m}$, we obtain the coupled equations for $c3$ and $c4$ in laboratory frame:
\begin{eqnarray}
    i\dot{c}_{3} &\approx&  (3\omega_{m}-\frac{6g^{2}}{\omega_{m}}) c_{3} + \frac{18\sqrt{3}g^{3}}{\omega_{m}^{2}} c_{4}, \\
    i\dot{c}_{4} &\approx&  (2\omega_{c}-\frac{27g^{2}}{\omega_{m}}) c_{4} + \frac{18\sqrt{3}g^{3}}{\omega_{m}^{2}} c_{3}.
\end{eqnarray}

 (2) For the approach of the Schrieffer-Wolff transformation, one should find a suitable unitary transformation to eliminate the direct photon-phonon coupling, and then we can get the effective Hamiltonian:
\begin{equation}
    H_{\rm eff}=e^{X}H_{s}e^{-X}\approx H_{0}+\frac{1}{2}[X,V]+\frac{1}{2}[X,[X,V]]+...,\label{32}
\end{equation}
where $H_{s}=H_{0}+V$ is the system Hamiltonian, $H_{0}=\omega_{c}a^{\dagger}a+\omega_{m}b^{\dagger}b$ is the unperturbed part, $V=g(a^{\dagger}+a)^{2}(b^{\dagger}+b)$ is the perturbed part, and the operator $X$ should be chosen to satisfy $[X,H_{0}]=-V$. Here, we choose 
\begin{equation}
    X=\frac{g}{4\omega_{m}}[(a^{\dagger})^{2}-2a^{2}+8a^{\dagger}a+4]b^{\dagger}-{\rm H.c.}.
\end{equation}
Under the frequency-matching condition ($2\omega_{c}=3\omega_{m}$), by keeping resonant terms only (in the rotating-wave approximation), we can calculate that
\begin{eqnarray}
\frac{1}{2}[X,V] &=& \frac{g^2}{8\omega_{m}}\left[ (a^{\dagger 2}-2a^{2}+8a^{\dagger}a+4)b - (a^{2}-2a^{\dagger 2}+8a^{\dagger}a+4)b, (a+a^{\dagger})^2 (b^{\dagger}+b) \right] \nonumber \\
&=& \frac{g^2}{8\omega_{m}} \Big\{ [2a^{\dagger 2}b, a^{2}b^{\dagger}] + [-2a^{2}b^{\dagger}, a^{\dagger 2}b] + [a^{\dagger 2}b^{\dagger}, a^{2}b] + [-a^{2}b, a^{\dagger 2}b^{\dagger}] \nonumber \\
&& + [(8a^{\dagger}a+4)b^{\dagger}, (2a^{\dagger}a+1)b] + [-(8a^{\dagger}a+4)b, (2a^{\dagger}a+1)b^{\dagger}] \Big\} \nonumber \\
&=&\frac{g^2}{2\omega_{m}}[a^{\dagger 2}a^2-2(a^{\dagger}a+1)b^{\dagger}b]+\frac{g^2}{4\omega_{m}}[-a^{\dagger2}a^2-2(a^{\dagger}a+1)b^{\dagger}b-2(a^{\dagger}a+1)]\nonumber \\
&&+\frac{g^2}{\omega_{m}}[-2(a^{\dagger}a+1)^2]\nonumber \\
&=&\frac{g^2}{4\omega_{m}}[a^{\dagger 2}a^2-2(a^{\dagger}a+1)(3b^{\dagger}b+4a^{\dagger}a+3)],
\end{eqnarray}
\begin{eqnarray}
\frac{1}{2}[X,[X,V]] &=& \frac{g^3}{32\omega_{m}^{2}} \left[ (a^{\dagger 2}-2a^{2}+8a^{\dagger}a+4)b - (a^{2}-2a^{\dagger 2}+8a^{\dagger}a+4)b, \right. \nonumber \\
&& \left. \left[ (a^{\dagger 2}-2a^{2}+8a^{\dagger}a+4)b - (a^{2}-2a^{\dagger 2}+8a^{\dagger}a+4)b, (a+a^{\dagger})^2 (b^{\dagger}+b) \right] \right] \nonumber \\
&=&-\frac{g^3}{4\omega_{m}^{2}}[a^{\dagger 2}a^2 a^{\dagger2}b^3+a^2a^{\dagger2}a^2 b^{\dagger3}]+\frac{g^3}{8\omega_{m}^2}[a^{\dagger 4}a^2 b^3+a^4a^{\dagger 2}b^{\dagger 3}]\nonumber \\
&&+\frac{g^3}{8\omega_{m}^2}[a^2 a^{\dagger 4}b^3+a^{\dagger2}a^4 b^{\dagger3}]+\frac{g^3}{2\omega_{m}^{2}}[(2a^{\dagger}a+1)^2a^2b^{\dagger3}+2a^{\dagger}a+1)^2a^{\dagger2}b^{3}] \nonumber \\
&&-\frac{g^3}{\omega_{m}^2}[(2a^{\dagger}a+1)a^2(2a^{\dagger}a+1)b^{\dagger3}+(2a^{\dagger}a+1)a^{\dagger 2}(2a^{\dagger}a+1)b^{3}]\nonumber \\
&&+\frac{g^3}{2\omega_{m}^2}[a^2(2a^{\dagger}a+1)^2 b^{\dagger3}+a^{\dagger 2}(2a^{\dagger}a+1)^2 b^{3}]\nonumber \\
&=&\frac{9g^3}{4\omega_{m}^2}\Big\{[(2a^{\dagger}a+1)a^{\dagger2}-a^{\dagger2}(2a^{\dagger}a+1)]b^3+ [a^{2}(2a^{\dagger}a+1)-(2a^{\dagger}a+1)a^{2}]b^{\dagger 3}\Big\} \nonumber \\
&=&\frac{9g^3}{\omega_{m}^2}(a^{\dagger 2}b^3+a^2 b^{\dagger 3}).\nonumber \\
\end{eqnarray}
Hence, one can obtain the effective Hamiltonian in Eq.~(\ref{21}).

\end{document}